\documentstyle[aps,preprint]{revtex}

\begin{document}
\draft
\title{Current transport in a superconducting superlattice system}
\author{Bor-Luen Huang and Chung-Yu Mou}
\address{
Department of Physics, National Tsing Hua University, Hsinchu 30043,
Taiwan}
\maketitle

\begin{abstract}
We investigate the effect of the superlattice structure on the single
particle transport along the c-axis of high temperature superconductors. In
particular, superlattice systems that consists of metals/insulators and
d-wave superconductors (NS/IS superlattice) are considered. We find that
for the NS superlattice in the large mass anisotropy limit of the metal, the
density of state in the low energy section is bulk d-wave like except that
the position of the quasi-particle peak can be considerably smaller than the
gap value, while for the IS superlattice, the quasi-particle peak remains at
the gap value. We also calculate the nonlinear differential conductance in
the planar junction measurement. It is found that the width of the Andreev
peak at zero-bias may be affected strongly by the superlattice structure,
specifically, it can be considerably reduced due to the destructive
interference of the Andreev reflections from all the superconductors. Such a
reduction in the width makes the Andreev peak resonant-like and has been
observed in a recent experiment.
\end{abstract}


\section{\protect\bigskip Introduction}

The artificial prepared superlattices have been an important system for
investigating the relevancy of periodic structure to the physical properties
of the bulk system. The idea is that if the dephasing length of the electron
extends over many periods, the superlattice structure becomes relevant.
Therefore, by engineering the periodic structure appropriately, one is able
to tune the effective parameters that electrons experience, thus it may
enable us to access more regions in the parameter space. In particular, one
may be able to engineer the density of state so that the transport
properties can be changed. Such systems have motivated a lot of studies both
experimentally and theoretically\cite{Ivchenko}. There are many outstanding
issues arising from the investigation of superlattice systems. For instance,
the giant magnetoresistance (MR) observed in magnetic superlattices, such as
Fe/Cr, has triggered serious investigation of the transport through
heterostructure of metal and ferromagnetic materials\cite{Oomi}.

The superlattice that has superconductors involved is another important
system. It was shown in an early study that certain superlattice structure
can close the energy gap of s-wave superconductors\cite{Hahn}, which is
interesting by itself and may have real applications in the future. The
interest in supperlatices with superconductors has been further boosted up
since the discovery of the high temperature superconductors (HTSC)\cite
{Martinez00}. The layered structure along the c-axis and particularly the
existence of closely packed $CuO_{2}$ planes makes HTSC a natural-occurring
superconducting superlattice system. A great body of work has thus been
devoted to investigate the c-axis dynamic properties during the past\cite
{caxis}. After the intense work for so many years, it now becomes clear that
in the overdoped region, the normal state of high Tc superconductors is
metallic like for both c-axis and in-plane directions\cite{cresisitivity}.
In particular, the splitting due to hopping between adjacent $CuO_{2}$
planes in a unit cell are seen directly in recent photoemission experiments 
\cite{Dessau}. While these works imply that single particle hopping
definitely occurs along the c direction, it is not clear how exactly one
should model the transport properties of the high Tc cuprates along the
c-axis in the superconducting phase and how the superlattice structure
affects the single particle transportation along the c direction.

The splitting observed in the photoemission experiments implies that
neighboring $CuO_{2}$ planes in a unit cell are strongly coupled by a large
hopping integral. Therefore, one can model it by a single superconducting
layer, which is coupled to the superconducting layers in adjacent unit cells
via an effective hopping integral $t_{eff}$. This is also consistent with
the fact that the coherence length along the c direction is about the width
of each superconducting layer. The problem is then how one should model
layers between adjacent superconducting layers. It was pointed in Ref.\cite
{Bulaevskii} that there exist two distinct limits, depending on the ratio\ $%
t_{eff}$ to $T_{c}$. In the limit when $t_{eff}$ \ is much smaller than $%
T_{c}$, the system behaves more like a NSNSN ...NSN (or a ISISISISI)
superlattice, i.e., a normal metal-superconductor (NS) or
insulator-superconductor (IS) superlattice. On the other hand, if $t_{eff}$
\ is close to $T_{c}$ in magnitude, different unit cells are coupled
strongly and the system behaves as a usual anisotropic superconductor. In
this work, we shall be mainly investigating the nonlinear transport in the
limit when $t_{eff}$ \ is much smaller than $T_{c}$. In order to model the
c-axis transport in HTSC, we shall consider several concrete models, for
instance, the NS and IS superlattice models as both of them can give rise
metallic behaviors in c and ab directions in the normal state. Another
possible modification is to consider a NS superlattice with a large mass
anisotropy (effective mass in the c-axis 
\mbox{$<$}%
\mbox{$<$}%
\ effective mass in ab directions) in the metal so that electrons
essentially hop along the c-axis in the NS superlattice model. Two
quantities will be calculated: (1) the density of state for single
particles, which will be useful for the transport measurement with high
resistance contact. (2) the differential conductance ($dI/dV$) for
measurements with low resistance contact , such as measurement made by
planar junctions. We shall include the d-wave nature and consider general
Fermi surface topology. Our results will be useful for artificial
superlattices such as the $YBa_{2}Cu_{3}O_{7-\delta }/\Pr
Ba_{2}Cu_{3}O_{7-\delta }$ superlattice system \cite{Martinez00}. At
phenomenological level, they also provide insights to the issue that to what
extent the c-axis of HTCS can be thought as a NS superlattice or a IS
superlattice, in particular, how the superlattice structure along the c-axis
affects the tunneling spectrum. Our results indicate that in the case for a
NS superlattice, if the mass anisotropy is small, the superlattice structure
will induce subgap structure in the density of state. For large mass
anisotropy in the metal, however, the bulk d-wave-like density of state is
reproduced with the quasi-particle peak shifted to a smaller value, in
contrast to the case when we model the c-axis as a IS superlattice where the
quasi-particle peak remains at the gap value. The NS superlattice model
provides a possible explanation of why some measurements of the gap size
along the c direction give smaller values\cite{Martinez00,Cheng}. We also
calculate the $dI/dV$ curve for measurements using planar junction or point
contact with large contact area along the c-direction. Two important
features are found to be due to the superlattice structure: (1) The width of
the Andreev peak at zero bias is considerably smaller than the gap due to
destructive interference of Andreev reflections from all the
superconductors. This gives a natural explanation to why in some old
measurement, the Andreev peak along the c-axis appeared to be so sharp and
was thought as zero-bias resonant conductance peak\cite{Sinha98}. (2) There
exist distinct oscillations in the region when $V$ is larger than the gap
size. This is also a result of interference from all in the interfaces. In
particular, the IS superlattice will provide a rising background so that
these oscillations gradually rises as $V$ increases. Both features are
observed in a recent measurement of Au/Bi2212 junctions near $T_{c}$\cite
{Chang}.

This paper is organized as follows: In Sec.II, the density of state is
calculated in the framework of Bogolubov de-Genne equation. We show that in
the large mass anisotropy limit of the NS superlattice, the usual d-wave
like behavior is reproduced. We also calculate the case when it is a SI
superlattice and briefly compare these results. In Sec.III, the non-linear $%
dI/dV$ curve for planar junctions is calculated and its relation to
experiments is discussed. Finally, we give a concluding remark in Sec.IV.

\section{Theoretical Formulation and density of state}

We shall start by investigating the density of state for a NS superlattice.
A NS superlattice with a spherical Fermi surface and s-wave for
superconductors was previously considered by Hahn\cite{Hahn} to model
transport along the c-axis for YBCO. To account for the c-axis transport of
high Tc cuprates, we shall extend it to include d-wave nature and use a more
general Fermi surface. In particular, during our calculation, we shall also
briefly address the effect due to a particular form for the c-axis hopping $%
t_{c}=-t_{\perp }\cos ^{2}(2\phi )$. This particular form of hopping is
suggested from the measurement of angle-resolved photoemission (ARPES) and
band theoretical calculation\cite{Dessau,Xiang}. Here $\phi $ is the angle $%
\tan ^{-1}(q_{y}/q_{x})$ (see Fig.1).

We shall first calculate the density of state. In the most general case, the
density of state is given by 
\begin{equation}
n(E)=\int \int \int \frac{dq_{x}dq_{y} d \kappa }{(2\pi )^{3}}\delta
(E-E(q_{x},q_{y},\kappa )),
\end{equation}
when $\kappa $ is the Block wavenumber along the c-axis. For given $E$, $%
q_{x}$, and $q_{y}$, if $\kappa (E,q_{x},q_{y})$ can be found, $n(E)$ can be
rewritten as 
\begin{equation}
n(E)=\int \int \frac{dq_{x}dq_{y}}{(2\pi )^{3}}n_{1}(E),  \label{dos1}
\end{equation}
where we have defined the one dimensional density of state by $%
n_{1}(E)\equiv \left| \frac{\partial \kappa }{\partial E}\right| $. In this
case, one needs to find $\kappa (E,q_{x},q_{y})$. For a NS superlattice, a
suitable framework for calculating $\kappa (E,q_{x},q_{y}) $ is the
Bogolubov de-Genne equation which can be written as

\begin{equation}
\left[ 
\begin{array}{rr}
\hat{{\rm h}} & \Delta \\ 
\Delta ^{\ast } & -\hat{{\rm h}}
\end{array}
\right] \left[ 
\begin{array}{c}
u({\bf r}) \\ 
\upsilon ({\bf r})
\end{array}
\right] \ =E\left[ 
\begin{array}{c}
u({\bf r}) \\ 
\upsilon ({\bf r})
\end{array}
\right] ,  \label{Bdg}
\end{equation}
where $\hat{{\rm h}}=\xi _{q}-\mu _{F}$, $\Delta =\Delta _{0}\cos (2\phi )$
in the superconducting part while in the metal part, we simply set $\Delta
_{0}=0$. We shall assume a general form for $\xi _{q}$ 
\begin{eqnarray*}
\xi _{q} &=&\frac{\hbar ^{2}}{2m^{\prime }}(q_{x}^{2}+q_{y}^{2})+\frac{\hbar
^{2}}{2m_{S}}q_{z}^{2}\text{ \ \ \ in the superconductor,} \\
&=&\frac{\hbar ^{2}}{2m}(q_{x}^{2}+q_{y}^{2})+\frac{\hbar ^{2}}{2m_{N}}%
q_{z}^{2}\text{ \ \ \ in the metal.}
\end{eqnarray*}
Since the system is translationally invariant in x and y directions, $q_{x}$
and $q_{y}$ are conserved so that we can write $u({\bf r})=u(z)\exp
^{i(q_{x}x+q_{y}y)}$ and $\upsilon ({\bf r})=\upsilon (z)\exp
^{i(q_{x}x+q_{y}y)}$ . Eqs.(\ref{Bdg}) then reduces to one dimensional
equations. In the metal part, $\Delta _{0}=0$, $q_{x}$ and $q_{y}$ are on
the Fermi surface so that we may parameterize $q_{x}=k_{F}\sin \theta \cos
\phi $ and $q_{y}=k_{F}\sin \theta \sin \phi $, where $k_{F}\equiv \sqrt{%
\frac{2m\mu _{F}}{\hbar ^{2}}}$ and $\theta $ is the azimuthal angle along
the c direction. In this parameterization, for a given energy $E$, the
particle ($+$) and hole ($-$) momentum become $p^{\pm }=\sqrt{\frac{2m_{S}}{%
\hbar ^{2}}[\mu _{F}(1-\frac{m}{m^{\prime }}\sin ^{2}\theta )\pm \sqrt{%
E^{2}-\Delta ^{2}}]}$ in the superconducting part, while $k^{\pm }=\sqrt{%
\frac{2m_{N}}{\hbar ^{2}}(\mu _{F}\cos ^{2}\theta \pm E)}$ in the metal.
Furthermore, Eq.(\ref{dos1}) becomes 
\begin{equation}
n(E)=\int \int \frac{k_{F}^{2}\sin \theta \cos \theta d\phi d\theta }{(2\pi
)^{3}}\frac{\partial \kappa }{\partial E}.  \label{dos2}
\end{equation}
It is important to note that the factor $1-m/m^{\prime }\sin ^{2}\theta $
has to be positive as it represents\ $(p^{\pm })^{2}$ at the Fermi surface
in the absence of $\Delta $. Therefore,{\em \ large }$m/m^{\prime }${\em ,
i.e., large mass anisotropy in the metal, restricts electrons to hop only
along the c-axis.}

The wavefunctions $u(z)$ and $\upsilon (z)$ have to to continuous at
boundaries. Since the system is periodic, it is sufficient to impose the
boundary conditions within a unit cell (see Fig. 1)

\begin{eqnarray}
u_{N}(0) &=&u_{S}(0),\upsilon _{N}(0)=\upsilon _{S}(0)  \nonumber \\
u_{N}^{\prime }(0) &=&u_{S}^{\prime }(0),\upsilon _{N}^{\prime }(0)=\upsilon
_{S}^{\prime }(0)  \nonumber \\
u_{S}(b) &=&\lambda u_{N}(-a),\upsilon _{S}(b)=\lambda \upsilon _{N}(-a) 
\nonumber \\
u_{S}^{\prime }(b) &=&\lambda u_{N}^{\prime }(-a),\upsilon _{S}^{\prime
}(b)=\lambda \upsilon _{N}^{\prime }(-a)
\end{eqnarray}
where the last two equations follow from the Bloch theorem with $\mid
\lambda \mid =1$. Following Ref.\cite{Gelder} and using the above boundary
conditions, one finds that $\lambda $ satisfies 
\begin{equation}
D_{0}+D_{1}\lambda +D_{2}\lambda ^{2}+D_{3}\lambda ^{3}+D_{4}\lambda ^{4}=0
\label{lamda}
\end{equation}
where $D_{0},D_{1},..,D_{4}$ are all real. Since$\mid \lambda \mid =1$, the
four roots to Eq.(\ref{lamda}) are in the form $\exp ^{\pm (i\theta _{1})}$
and $\exp ^{\pm (i\theta _{2})}$. This implies that $D_{0}=D_{4},D_{1}=D_{3}$%
. Thus the solutions to Eq.(\ref{lamda}) is completely determined by the
ratio $\frac{D_{1}}{D_{0}}$ and $\frac{D_{2}}{D_{0}}$. 
\begin{equation}
\cos (\kappa \,d)=\frac{1}{4}\left[ -\frac{D_{1}}{D_{0}}\pm \sqrt{\left( 
\frac{D_{1}}{D_{0}}\right) ^{2}-4\frac{D_{2}}{D_{0}}+8}\right] ,\newline
\label{cos}
\end{equation}
where we have expressed $\lambda $ by $\exp (i\kappa d)$ with $d=a+b$. We
shall see that the positive sign is for the particle excitation, while the
negative sign is for the hole excitation. It is convenient to measure the
length by $\xi _{N}\equiv \hbar ^{2}k_{F}/\sqrt{mm_{N}\Delta _{0}}$ and
energy by $\Delta _{0}$ ($\varepsilon \equiv E/\Delta _{0}$). Denoting $%
A\equiv a/\xi _{N}$, $B\equiv b/\xi _{N}$ and $\epsilon _{F}\equiv \mu
_{F}/\Delta _{0}$, after some algebra and using the Andreev approximation
\cite{Gelder}, we find that for a fixed $\phi $ and $\theta $ 
\begin{eqnarray}
\frac{D_{1}}{D_{0}} &=&-4\left[ \cos (2\varepsilon _{F}AX)\cos (2\varepsilon
_{F}\gamma BX_{m})\Gamma _{1}-\sin (2\varepsilon _{F}AX)\sin (2\varepsilon
_{F}\gamma BX_{m})\Gamma _{2}\right] ,  \label{D1} \\
\frac{D_{2}}{D_{0}} &=&4\Gamma _{2}^{2}+(g^{2}-1)\left[ 1+\cos (4\varepsilon
_{F}AX)+\cos (4\varepsilon _{F}\gamma BX_{m})-2\Gamma _{3}\right]   \nonumber
\\
&&+(g^{2}+1)\cos (4\varepsilon _{F}AX)\cos (4\varepsilon _{F}\gamma
BX_{m})-2g\sin (4\varepsilon _{F}AX)\sin (4\varepsilon _{F}\gamma BX_{m}).
\label{D2}
\end{eqnarray}
Here $X=\cos \theta $, $\gamma =\sqrt{m_{S}/m_{N}}$, $X_{m}=\sqrt{%
1-(m/m^{\prime })\sin ^{2}\theta }$ and $g\equiv (\cos \theta /\gamma
X_{m}+\gamma X_{m}/\cos \theta )/2$. For $\epsilon ^{2}>\cos ^{2}2\phi $, $%
\Gamma _{i}$ are given by 
\begin{eqnarray}
\Gamma _{1} &=&\cos \left( \frac{\varepsilon A}{X}\right) \cos \left( \frac{%
\varepsilon _{\phi }B}{X_{m}}\right) -g\frac{\varepsilon }{\varepsilon
_{\phi }}\sin \left( \frac{\varepsilon A}{X}\right) \sin \left( \frac{%
\varepsilon _{\phi }B}{X_{m}}\right) ,  \label{G1} \\
\Gamma _{2} &=&g\cos \left( \frac{\varepsilon A}{X}\right) \cos \left( \frac{%
\varepsilon _{\phi }B}{X_{m}}\right) -\frac{\varepsilon }{\varepsilon _{\phi
}}\sin \left( \frac{\varepsilon A}{X}\right) \sin \left( \frac{\varepsilon
_{\phi }B}{X_{m}}\right) ,  \label{G2} \\
\Gamma _{3} &=&\cos ^{2}\left( \frac{\varepsilon _{\phi }B}{X_{m}}\right) %
\left[ 1+\cos (4\varepsilon _{F}AX)\right] +\cos ^{2}\left( \frac{\epsilon A%
}{X}\right) \left[ 1+\cos (4\varepsilon _{F\ }\gamma BX_{m})\right] ,
\label{G3}
\end{eqnarray}
where $\varepsilon _{\phi }=\sqrt{\varepsilon ^{2}-\cos ^{2}2\phi }$; while
for $\varepsilon ^{2}<\cos ^{2}2\phi $, one simply replaces $\cos $($\sin $
) by $\cosh $ ($\sinh $), and $\varepsilon _{\phi }$ by$\sqrt{\cos ^{2}2\phi
-\varepsilon ^{2}}$.

{\em Homogenous case} \ \ This is the case when $m=m^{\prime }$ and $%
m_{N}=m_{S}$. Therefore, $g=1$. We have $\Gamma _{1}=\Gamma _{2}\equiv
\Gamma (\varepsilon )$, as a result, we find that $\cos (\kappa \,d)=\Gamma
\cos [2\varepsilon _{F}(A+B)X]\pm \sqrt{1-\Gamma ^{2}}\sin [2\varepsilon
_{F}(A+B)X]$, recovering results obtained in Ref.\cite{Gelder} with $\frac{%
k^{+}+k^{-}}{2}$ and $\frac{p^{+}+p^{-}}{2}$ being approximated by $%
2\varepsilon _{F}(A+B)X$. It is instructive to check the case of $\Delta
_{0}=0$. In this case, we have $p^{\pm }=k^{\pm }$ and thus $\Gamma =1.$ We
find that $\kappa =k^{\pm }$. Therefore, $\pm $ represents particle and hole
channels respectively. For the homogeneous case, since $\Gamma ^{2}\leq 1$,
we may write $\Gamma =\sin \alpha (\varepsilon )$. It is then easy to see
that $n_{1}(\varepsilon )=\frac{1}{d}\left| \pm \frac{\partial \alpha }{%
\partial \varepsilon }\right| $ are the same for particles and holes. As
shown in Fig. 2(a), if we use negative $\epsilon $ to represent the hole,
the one dimensional density of state is an even function of $\varepsilon $.
Fig. 2(b) shows the result when $m_{S}\neq m_{N}$, it is seen that there is
no particle-hole symmetry any more. Note that the values of $A$ and $B$ are
chosen to be close the data for YBCO: $\xi _{N}\approx \xi _{ab}\approx
1.5nm $, $a\approx 0.85nm$, and $b\approx 0.38nm$. It is also instructive to
check that the NS superlattice model also includes the limit when $t_{eff}$
\ is close to $T_{c}$ in magnitude as one can simply set $A=0$, $B=1$ in the
homogeneous case. In this case, $\Gamma $ is simply $\cosh \sqrt{\varepsilon
^{2}-\cos ^{2}2\phi }$ for $\varepsilon ^{2}\geq \cos ^{2}2\phi $ and $\cos 
\sqrt{\cos ^{2}2\phi -\varepsilon ^{2}}$ for $\varepsilon ^{2}<\cos
^{2}2\phi $. Thus $n_{1}(\varepsilon )=\frac{1}{d}\frac{\varepsilon }{\sqrt{%
\varepsilon ^{2}-\cos ^{2}2\phi }}$ for $\varepsilon ^{2}\geq \cos ^{2}2\phi 
$, which reproduces the usual d-wave density of state after integration over 
$\phi $, see Fig.3(a).

Fig. 3(b) shows a case when $A$ becomes nonzero. It is obvious that the
introduction of any small N section (i.e., small $A$) moves the
quasi-particle peak into the subgap region while\ leaving in a peak-like
structure at $E=\Delta _{0}$. Fig. 3(c) shows a more realistic case for $%
A=0.57$ and $B=0.21$ (homogeneous). The structure at $E=\Delta _{0}$ becomes
too small to be observed.

{\em Large mass anisotropy limit} \ We now discuss the effect of increasing
the ratio$\ m/m^{\prime }$. A possible candidate to describe the high Tc
cuprates is the limit when $m/m^{\prime }$ becomes so large that electrons
essentially can only hop along the c-axis between CuO$_{2}$ planes in
different unit cells. As we pointed it out already that large $m/m^{\prime }$
restricts $\theta $ to be small. For infinite $m/m^{\prime }$, we can simply
set $X=\cos \theta =1$ and perform the following reduced integral 
\begin{equation}
n_{r}(E)=\left. \int \frac{k_{F}^{2}d\phi }{(2\pi )^{3}}\frac{\partial
\kappa }{\partial E}\right| _{X=1}.  \label{inf}
\end{equation}
If the superconducting part is s-wave, $n_{r}(E)$ is $\left. \frac{k_{F}^{2}%
}{(2\pi )^{2}}\frac{\partial \kappa }{\partial E}\right| _{X=1}$. Consider
the case when\ $m_{S}=m_{N}$, it is easy to see that $X_{\theta }=g=1$. We
find that $\Gamma _{1}=\Gamma _{2}\equiv \Gamma (\varepsilon )$ and $\cos
(\kappa \pm 2\varepsilon _{F}(A+B)X)d=\Gamma (\varepsilon )$. The point is
that the one dimensional density of state $\partial \kappa /\partial
\varepsilon $ starts from a gap whose size is determined by the root to $%
\Gamma (\varepsilon _{0})=1$ for $\varepsilon _{0}\leq 1$. Near $\varepsilon
_{0}^{+}$, one has 
\begin{equation}
|\frac{\partial \kappa }{\partial \varepsilon }|=\frac{1}{d}\frac{1}{\sqrt{%
1-\Gamma ^{2}}}|\frac{\partial \Gamma }{\partial \varepsilon }|\approx \frac{%
1}{d}\sqrt{\frac{\Gamma ^{\prime }(\varepsilon _{0})}{2}}\frac{1}{\sqrt{%
\varepsilon -\varepsilon _{0}}},  \label{inf2}
\end{equation}
which has a BCS-like square root singularity but at smaller energy ($%
\varepsilon _{0}<1$). Therefore, in the lower energy section, we obtain a
bulk s-wave like density of state with quasi-particle peak at smaller
energy. For high energy sections, near the zone boundary of $\kappa $, the
BCS-like square root singularity will repeat again. For d-wave
superconductors, Eq.(\ref{inf}) is replaced by 
\begin{equation}
|\frac{\partial \kappa }{\partial \varepsilon }|=\frac{1}{d}\frac{1}{\sqrt{%
1-\Gamma ^{2}}}|\frac{\partial \Gamma }{\partial \varepsilon }|\approx \sqrt{%
\frac{\Gamma ^{\prime }(\varepsilon _{0})}{2}}\frac{|\cos 2\phi |}{\sqrt{%
\varepsilon -\varepsilon _{0}|\cos 2\phi |}},  \label{inf3}
\end{equation}
where $\varepsilon _{0}$ depends on $\phi $. Thus, the BCS-like square root
singularity is preserved for a fixed $\phi $. In Fig.4, we show the result
of direct integration over $\phi $ for different $m_{S}/m_{N}$ ratios. It is
clear that at low energy section, $n_{r}(E)$ behaves similar to the bulk
d-wave density of state, except that the quasi-particle peak moves to
smaller energy. If we take $\Delta _{0}\approx 30-40$ ${\rm meV}$, the
quasi-particle peak is around $8-11$ {\rm meV }which is close to what
experiments have seen\cite{Martinez00}. Note that another difference between
the above results and the true bulk d-wave density of state may lie at high
energy section. In general, the superlattice structure introduces coupling
among the wavevector of the electron and the reciprocal lattice vectors of
the superlattice, i.e., $(0,0,2n\pi /d)$ ($\equiv {\bf Q}$). The coupling
strength depends on the magnitude of the Fourier component $\Delta _{0}({\bf %
Q})$. If \ the superlattice periodicity is good, $\Delta _{0}({\bf Q})$ will
not be small. The superlattice structure will force $n_{r}(E)$ to have
similar d-wave like structure whenever the zone boundary ($\cos \kappa d=\pm
1$) is encountered in high energy sectors. However, if the system is not
large enough along the c direction or the dephasing length of the electron
is short, one may expect that $\Delta _{0}({\bf Q})$ is small and hence the
repeated structure in high energy section will not appear.

We now address the effect due to the anisotropy of the c-axis hopping. In
particular, we consider $t_{c}=-t_{\perp }\cos ^{2}(2\phi )$. As it is clear
from the above calculations and also the following calculation for the IS
superlattice, the density of state depends only on the ratio of $m_{S}$ to $%
m_{N}$ (defined as $\gamma ^{2}$). Obviously, if both the
intra-superconducting cells and inter-superconducting cells c-axis hoppings
follow the same form. There will be no effect at all. However, new features
may arise if \ they do not follow the same form. In particular, we find that
if either $m_{S}$ or $m_{N}$ does not depend on $\phi $, a second peak could
arise inside the subgap region. In views of past experimental findings, this
seems to be unlikely. Therefore, we shall not consider such possibility in
the following.

{\em IS superlattice \ }As a comparison, we now consider the case when the
materials between CuO$_{2}$ planes in different unit cells are modeled by
insulators. For this purpose, we introduce a large potential $V$ to every
metal cell so that the metal cells effectively become insulators. We shall
consider the simplest case when $m=m^{\prime }=m_{N}=m_{S}$. When the
superconducting gap vanishes, this model reduces to the standard
Kronig-Penny, therefore, it represents a natural generalization of NIN
superlattice. In this case, $k^{\pm }$ becomes purely imaginary. Following
the same procedure, we find that Eqs. (\ref{D1}) and (\ref{D2}) are
essentially the same except that $g$ is now defined by $i(\alpha +1/\alpha
)/2$ with $\alpha =\sqrt{\mu _{F}\cos ^{2}\theta /(V-\mu _{F}\cos ^{2}\theta
)}$ and $\cos (2\varepsilon _{F}AX)$ and $\sin (2\varepsilon _{F}AX)$ being
replaced by $\cosh (2\varepsilon _{F}AX/\alpha )$ and $\sinh (2\varepsilon
_{F}AX/\alpha )$. The new $\Gamma _{i}$ are given by (for $\epsilon
^{2}>\cos ^{2}2\phi $) 
\begin{eqnarray}
\Gamma _{1} &=&\cosh \left( \frac{\varepsilon A\alpha }{X}\right) \cos
\left( \frac{\varepsilon _{\phi }B}{X}\right) +ig\frac{\varepsilon }{%
\varepsilon _{\phi }}\sinh \left( \frac{\varepsilon A\alpha }{X}\right) \sin
\left( \frac{\varepsilon _{\phi }B}{X}\right) , \\
\Gamma _{2} &=&-ig\cosh \left( \frac{\varepsilon A\alpha }{X}\right) \cos
\left( \frac{\varepsilon _{\phi }B}{X}\right) -\frac{\varepsilon }{%
\varepsilon _{\phi }}\sinh \left( \frac{\varepsilon A\alpha }{X}\right) \sin
\left( \frac{\varepsilon _{\phi }B}{X}\right) , \\
\Gamma _{3} &=&\cos ^{2}\left( \frac{\varepsilon _{\phi }B}{X}\right) \left[
1+\cosh (\frac{4\varepsilon _{F}AX}{\alpha })\right] +\cosh ^{2}\left( \frac{%
\epsilon A\alpha }{X}\right) \left[ 1+\cos (4\varepsilon _{F\ }BX)\right] ,
\end{eqnarray}
while again for $\varepsilon ^{2}<\cos ^{2}2\phi $, one replaces $\cos $($%
\sin $ ) by $\cosh $ ($\sinh $), and $\varepsilon _{\phi }$ by$\sqrt{\cos
^{2}2\phi -\varepsilon ^{2}}$. In Fig.5, we show a \ typical result for the
three dimensional density of state for the SI superlattice. Because for
energy below the gap, $\varepsilon ^{2}<\cos ^{2}2\phi $, quasi particles
and quasi holes are both evanescent, the quasi-particle peaks now move to
the gap value (except that there is a slight asymmetry in particles and
holes). Technically, this is due to that for $\varepsilon ^{2}<\cos
^{2}2\phi $, all factors in $\Gamma _{i}$ are not oscillatory and thus the
right hand side in Eq.(\ref{cos}) is greater than one, without any
propagating solutions. This is similar to the case for a bulk d-wave
superconductor as we analyzed for Fig.3(a), so one gets quasi-particle peaks
right at the gap value.

The dramatic difference in the positions of the quasi-particle peaks for the
NS and IS superlattices offers a useful check on both models. For artificial
made NS superlattices, the reduction in the position of \ the quasi-particle
peak has been observed in Ref.\cite{Martinez00}. However, for the c-axis
measurements on real high Tc materials, the results are controversial and
remains to be clarified in the future\cite{covington00}.

\section{Nonlinear differential conductance}

We now investigate the differential conductance ($dI/dV$) for the
measurement with low resistance contact, such as measurements made by planar
junctions or by point contact with large contact area. The schematic setup
is shown in Fig.6. In principle, at first and last interfaces that connect
with the electrodes, there may exist insulating layers. This will be
included as interfacial scattering potentials. The most important feature
coming out from this type of measurement, in comparison to the high contact
resistance measurement, is the possible exhibition of the excess current due
to the Andreev reflection (AR). However, for a long time, the AR along the c
direction was not observed until quite recently Andreev peaks with reduced
widths are observed in Au/Bi2212 junctions near $T_{c}$\cite{Chang}.

In this section, we shall carry out a theoretical calculation of the $dI/dV$
curves based on the NS/IS superlattice models. We shall show that the width
of the Andreev peak in such systems is often reduced due to the superlattice
structure. A similar problem in which the superconductor in each cell is
replaced by another metal (N$^{\prime }$) was analyzed in Ref.\cite{Ning}
using a full quantum mechanic approach. Here we shall approximate the
voltages in each layer (either in N or S) by constants such that the slope
of the voltage, thus the electric field, is fixed. This is exact in the
superconducting cell, however, in the metal cell the approximation is valid
only if the electric field is weak (specifically, $eVa/nd$ has to be smaller
than $\varepsilon _{F}$). To obtain the differential conductance, we need to
match the quasi-particle wave functions at boundaries: $z=nd$ and $z=nd+b$.
At the boundary $z=nd$, the quasi-particle wave function for the normal
metal is

\begin{eqnarray}
\psi (z=nd)\equiv \left( 
\begin{array}{c}
u(z=nd) \\ 
\upsilon (z=nd)
\end{array}
\right) = &&A_{n}\left( 
\begin{array}{c}
1 \\ 
0
\end{array}
\right) e^{ik_{n}^{+}nD}+B_{n}\left( 
\begin{array}{c}
1 \\ 
0
\end{array}
\right) e^{-ik_{n}^{+}nD}  \nonumber \\
+ &&C_{n}\left( 
\begin{array}{c}
0 \\ 
1
\end{array}
\right) e^{ik_{n}^{-}nD}+D_{n}\left( 
\begin{array}{c}
0 \\ 
1
\end{array}
\right) e^{-ik_{n}^{-}nD},
\end{eqnarray}
where we have measured energy in unit of $\Delta _{0}$ and lengths by $\xi
_{N}$, hence $D=d/\xi _{N}$ and $k_{n}^{\pm }=\sqrt{\cos ^{2}\theta +(\pm
\varepsilon -V_{n}^{N})/\varepsilon _{F}}$ with $V_{n}^{N}=\frac{V}{\Delta
_{0}}(1-n/N_{0})$ and $N_{0}$ being the total number of layers. For the
superconducting region at the right hand side of $z=nd$, the wavefunction is

\begin{eqnarray}
\phi (z=nd)= &&E_{n+1}\left( 
\begin{array}{c}
u_{0} \\ 
\upsilon _{0}
\end{array}
\right) e^{ip_{n+1}^{+}nD}+F_{n+1}\left( 
\begin{array}{c}
u_{0} \\ 
\upsilon _{0}
\end{array}
\right) e^{-ip_{n+1}^{+}nD}  \nonumber \\
+ &&G_{n+1}\left( 
\begin{array}{c}
\upsilon _{0} \\ 
u_{0}
\end{array}
\right) e^{ip_{n+1}^{-}nD}+H_{n+1}\left( 
\begin{array}{c}
\upsilon _{0} \\ 
u_{0}
\end{array}
\right) e^{-ip_{n+1}^{-}nD}.
\end{eqnarray}
Here the wavevector $p_{n+1}^{\pm }$ is measured by $k_{F}$ and is given by $%
\sqrt{\gamma _{1}(1-\gamma _{2}\sin ^{2}\theta )+\gamma _{1}/\varepsilon
_{F}(\pm \varepsilon _{\phi }-V_{n+1}^{S})]}$ with $\gamma _{1}\equiv \frac{%
m_{S}}{m_{N}}$ and $\gamma _{2}\equiv \frac{m}{m^{\prime }}$, $V_{n+1}^{S}=%
\frac{V}{\Delta _{0}}[1-(nD+B)/N_{0}D]$ (measured by $k_{F}$), and the
coherent factors $u_{0}^{2}=(1+\varepsilon _{\phi }/\varepsilon )/2$ and $%
\upsilon _{0}^{2}=(1-\varepsilon _{\phi }/\varepsilon )/2$. The boundary
conditions at $z=nd$ are

\[
\psi (nd)=\phi (nd), 
\]

\begin{equation}
\frac{m_{N}}{m_{S}}\frac{\partial }{\partial z}\phi (nd)-\frac{\partial }{%
\partial z}\psi (nd)=\frac{2m_{N}H}{\hbar ^{2}}\phi (nd),
\end{equation}
where we have introduced a delta potential with strength $H$ at this
interface\cite{BTK}. This provides a relation between $%
A_{n},B_{n},C_{n},D_{n}$ and $E_{n+1},F_{n+1},G_{n+1},H_{n+1}$, which is
conveniently expressed by introducing a transfer matrix $M_{1}$

\begin{equation}
M_{1}\left( 
\begin{array}{c}
E_{n+1} \\ 
F_{n+1} \\ 
G_{n+1} \\ 
H_{n+1}
\end{array}
\right) =\left( 
\begin{array}{c}
A_{n} \\ 
B_{n} \\ 
C_{n} \\ 
D_{n}
\end{array}
\right) .
\end{equation}
Here the full expression of $M_{1}$ is given by 
\begin{eqnarray}
&&\left( 
\begin{array}{cc}
\frac{u_{0}}{2}[1+i\frac{2Z}{k_{n}^{+}}+\frac{p_{n+1}^{+}}{\gamma
_{1}k_{n}^{+}}]e^{i2\epsilon _{F}[p_{n+1}^{+}-k_{n}^{+}]nD} & \frac{u_{0}}{2}%
[1+i\frac{2Z}{k_{n}^{+}}-\frac{p_{n+1}^{+}}{\gamma _{1}k_{n}^{+}}%
]e^{-i2\epsilon _{F}[p_{n+1}^{+}+k_{n}^{+}]nD} \\ 
\frac{u_{0}}{2}[1-i\frac{2Z}{k_{n}^{+}}-\frac{p_{n+1}^{+}}{\gamma
_{1}k_{n}^{+}}]e^{i2\epsilon _{F}[p_{n+1}^{+}+k_{n}^{+}]nD} & \frac{u_{0}}{2}%
[1-i\frac{2Z}{k_{n}^{+}}+\frac{p_{n+1}^{+}}{\gamma _{1}k_{n}^{+}}%
]e^{-i2\epsilon _{F}[p_{n+1}^{+}-k_{n}^{+}]nD} \\ 
\frac{\upsilon _{0}}{2}[1+i\frac{2Z}{k_{n}^{-}}+\frac{p_{n+1}^{+}}{\gamma
_{1}k_{n}^{-}}]e^{i2\epsilon _{F}[p_{n+1}^{+}-k_{n}^{-}]nD} & \frac{\upsilon
_{0}}{2}[1+i\frac{2Z}{k_{n}^{-}}-\frac{p_{n+1}^{+}}{\gamma _{1}k_{n}^{-}}%
]e^{-i2\epsilon _{F}[p_{n+1}^{+}+k_{n}^{-}]nD} \\ 
\frac{\upsilon _{0}}{2}[1-i\frac{2Z}{k_{n}^{-}}-\frac{p_{n+1}^{+}}{\gamma
_{1}k_{n}^{-}}]e^{i2\epsilon _{F}[p_{n+1}^{+}+k_{n}^{-}]nD} & \frac{\upsilon
_{0}}{2}[1-i\frac{2Z}{k_{n}^{-}}+\frac{p_{n+1}^{+}}{\gamma _{1}k_{n}^{-}}%
]e^{-i2\epsilon _{F}[p_{n+1}^{+}-k_{n}^{-}]nD}
\end{array}
\right.   \nonumber \\
&&\left. 
\begin{array}{cc}
\frac{\upsilon _{0}}{2}[1+i\frac{2Z}{k_{n}^{+}}+\frac{p_{n+1}^{-}}{\gamma
_{1}k_{n}^{+}}]e^{i2\epsilon _{F}[p_{n+1}^{-}-k_{n}^{+}]nD} & \frac{\upsilon
_{0}}{2}[1+i\frac{2Z}{k_{n}^{+}}-\frac{p_{n+1}^{-}}{\gamma _{1}k_{n}^{+}}%
]e^{-i2\epsilon _{F}[p_{n+1}^{-}+k_{n}^{+}]nD} \\ 
\frac{\upsilon _{0}}{2}[1-i\frac{2Z}{k_{n}^{+}}-\frac{p_{n+1}^{-}}{\gamma
_{1}k_{n}^{+}}]e^{i2\epsilon _{F}[p_{n+1}^{-}+k_{n}^{+}]nD} & \frac{\upsilon
_{0}}{2}[1-i\frac{2Z}{k_{n}^{+}}+\frac{p_{n+1}^{-}}{\gamma _{1}k_{n}^{+}}%
]e^{-i2\epsilon _{F}[p_{n+1}^{-}-k_{n}^{+}]nD} \\ 
\frac{u_{0}}{2}[1+i\frac{2Z}{k_{n}^{-}}+\frac{p_{n+1}^{-}}{\gamma
_{1}k_{n}^{-}}]e^{i2\epsilon _{F}[p_{n+1}^{-}-k_{n}^{-}]nD} & \frac{u_{0}}{2}%
[1+i\frac{2Z}{k_{n}^{-}}-\frac{p_{n+1}^{-}}{\gamma _{1}k_{n}^{-}}%
]e^{-i2\epsilon _{F}[p_{n+1}^{-}+k_{n}^{-}]nD} \\ 
\frac{u_{0}}{2}[1-i\frac{2Z}{k_{n}^{-}}-\frac{p_{n+1}^{-}}{\gamma
_{1}k_{n}^{-}}]e^{i2\epsilon _{F}[p_{n+1}^{-}+k_{n}^{-}]nD} & \frac{u_{0}}{2}%
[1-i\frac{2Z}{k_{n}^{-}}+\frac{p_{n+1}^{-}}{\gamma _{1}k_{n}^{-}}%
]e^{-i2\epsilon _{F}[p_{n+1}^{-}-k_{n}^{-}]nD}
\end{array}
\right) ,
\end{eqnarray}
with $Z$ ($\equiv \frac{Hm_{N}}{\hbar ^{2}k_{F}}$) characterizing the
interface potential. Note that in principle different layers may have
different $Z$. The above formalism is for the NS superlattice. To model a IS
superlattice, one adds a large potential $V$ to the metal cell so that $%
k_{n}^{\pm }$ become evanescent and are replaced by $\sqrt{\cos ^{2}\theta +%
\frac{1}{\varepsilon _{F}}(\pm \varepsilon -V_{n}^{N}-\frac{V}{\Delta _{0}})}
$.

Similar boundary conditions are also imposed at $z=nd+b$, and yield the
relation 
\begin{equation}
M_{2}\left( 
\begin{array}{c}
E_{n+1} \\ 
F_{n+1} \\ 
G_{n+1} \\ 
H_{n+1}
\end{array}
\right) =\left( 
\begin{array}{c}
A_{n+1} \\ 
B_{n+1} \\ 
C_{n+1} \\ 
D_{n+1}
\end{array}
\right)
\end{equation}
where $M_{2}$ can be obtained by $M_{1}$ simply by the following changes: $%
nD\rightarrow nD+B$, $k_{n}^{\pm }\rightarrow k_{n+1}^{\pm }$ and $%
H\rightarrow -H$. Thus the effective transfer matrix that connects the nth
cell and (n+1)th cell has the form

\begin{equation}
M_{n\rightarrow n+1}\left( 
\begin{array}{c}
A_{n} \\ 
B_{n} \\ 
C_{n} \\ 
D_{n}
\end{array}
\right) =\left( 
\begin{array}{c}
A_{n+1} \\ 
B_{n+1} \\ 
C_{n+1} \\ 
D_{n+1}
\end{array}
\right) ,
\end{equation}
where $M_{n\rightarrow n+1}=M_{2}\cdot M_{1}^{-1}$. Note that coherence in
superconducting region is assumed so that no phase memory is lost when
quasi-particles tunnel from one superconducting cell to another one. The
transfer matrix of the whole system for $N_{0}$-layers, $\hat{T}_{N_{0}}$,
is thus obtained by

\begin{equation}
\hat{T}_{N_{0}}=\prod_{n=0}^{N_{0}-1}M_{n\rightarrow n+1},
\end{equation}
so the coefficients in the N$_{0}$-th cell $%
A_{N_{0}},B_{N_{0}},C_{N_{0}},D_{N_{0}}$ are connected with those in first
cell $A_{0},B_{0},C_{0},D_{0}$ by the relation

\begin{equation}
\hat{T}_{N_{0}}\left( 
\begin{array}{c}
A_{0} \\ 
B_{0} \\ 
C_{0} \\ 
D_{0}
\end{array}
\right) =\left( 
\begin{array}{c}
A_{N_{0}} \\ 
B_{N_{0}} \\ 
C_{N_{0}} \\ 
D_{N_{0}}
\end{array}
\right) .
\end{equation}
We shall further impose the boundary conditions near electrodes. Firstly,
because electron can only flow into the system from the high voltage
electrode, we require

\begin{equation}
A_{0}=1,D_{0}=0.  \label{eq1}
\end{equation}
Secondly, in the low voltage electrode, we impose the condition

\begin{equation}
B_{N_{0}}=C_{N_{0}}=0,  \label{eq2}
\end{equation}
so that no particle (both hole and electron) flows into the system from the
lowest voltage side. Eqs. (\ref{eq1}) and(\ref{eq2}) are then sufficient to
determine the amplitude $B_{0}$ and $C_{0}$ in terms components of $%
T_{N_{0}} $, we find that

\begin{eqnarray}
B_{0}
&=&(T_{N_{0}}^{31}T_{N_{0}}^{23}-T_{N_{0}}^{21}T_{N_{0}}^{33})/(T_{N_{0}}^{22}T_{N_{0}}^{33}-T_{N_{0}}^{32}T_{N_{0}}^{23})
\nonumber \\
C_{0}
&=&(T_{N_{0}}^{31}T_{N_{0}}^{22}-T_{N_{0}}^{21}T_{N_{0}}^{32})/(T_{N_{0}}^{23}T_{N_{0}}^{32}-T_{N_{0}}^{22}T_{N_{0}}^{33}).
\label{BC}
\end{eqnarray}
To obtain the differential conductance in terms of the above coefficients,
we apply the same argument used in Ref.\cite{BTK} in which one treats the
superlattice between electrodes as a scattering center. We obtain the same
expression for the differential conductance

\begin{equation}
dI/dV=2N(E_{F})ev_{F}{\cal A}(1-\mid B_{0}(eV)\mid ^{2}+\mid C_{0}(eV)\mid
^{2}),  \label{dIdV}
\end{equation}
where ${\cal A}$ is the area for the junction.

To illustrate the behavior of the differential conductance, it is useful to
consider the simplest case when $N_{0}=1$, i.e., a NSN structure. This will
be the more realistic configuration for measurement on what is often
referred as the NS junction in the literature\cite{BTK}. We shall consider
the case when $m=m^{\prime }=m_{N}=m_{S}$ and $\varepsilon _{F}>>1$ so that
one can apply the so-called Andreev approximation in which one approximates $%
k_{n}^{\pm }$ and $p_{n}^{\pm }$ by $1$ ($\theta =0$ and $n=0$), i.e., the
wavevectors along c-axis are approximated by $k_{F}$. This is the
approximation used to derive the differential conductance of the NS junction
in the BTK theory\cite{BTK}. Nevertheless, it turns out that for a NSN
junction, such approximation leads to a result with zero Andreev current. A
closer examination shows that in the large $\varepsilon _{F}$, one has to
keep the expansion of both $k_{n}^{\pm }$ and $p_{n}^{\pm }$ to the first
order in $(\pm \varepsilon -V_{n}^{N})/\varepsilon _{F}$ or $(\pm
\varepsilon _{\phi }-V_{n+1}^{S})/\varepsilon _{F}$ when computing the phase
terms. We find that such manipulation leads to 
\begin{eqnarray}
B_{0} &=&0  \nonumber \\
C_{0} &=&\frac{-1+e^{2\varepsilon _{\phi }Bi}}{-u_{0}^{2}+\upsilon
_{0}^{2}e^{2\varepsilon _{\phi }Bi}}u_{0}\upsilon _{0}.
\end{eqnarray}
Therefore, for $\varepsilon ^{2}<\cos ^{2}2\phi $, $\varepsilon _{\phi }$ is
purely imaginary and hence in the limit of large $B$, $C_{0}$ approaches $%
-\upsilon _{0}/u_{0}$, recovering the standard BTK result. For $\varepsilon
^{2}\geq \cos ^{2}2\phi $, $\varepsilon _{\phi }$ is positive and $C_{0}$
oscillates with periods, determined only by $B$. The oscillation of $C_{0}$
was actually observed in Ref.\cite{BTK1}. It is a result of interference of
Andreev reflection from both interfaces (left and right interfaces) of the
superconductors. In Fig.7(a), we show the $dI/dV$ curves of directional
tunneling ($\phi =0$ and without using the Andreev approximation) for
different $B$ with $A$ being fixed at one. One sees that the period of
oscillation decreases for increasing $B$, in consistent with the above
analysis.

As $N_{0}$ exceeds one, \ the lengthscale $A$ also participates in
determining the oscillations. In this case, the Andreev approximation always
yields $B_{0}=0$ when $Z=0$ because $T_{N_{0}}^{21}$ and $T_{N_{0}}^{32}$
are always zero in this approximation. The total conductance is thus
determined by $C_{0}$ given by $-T_{N_{0}}^{31}/T_{N_{0}}^{33}$. Fig. 7(b)
shows the calculated differential conductance for the directional tunneling
(electrons incident perpendicular to the interface) of a NS superlattice by
using the exact expressions of the transfer matrix. As a demonstration, here
we consider the homogeneous case with the total number of layers being ten.
One sees that a repeated main Andreev peak is already seen around $%
\varepsilon \approx 4$, resembling the repeated structure when crossing the
zone boundary in infinite systems. Note that these repeated structures may
not be exhibited in real experiments as the dephasing length could easily
get shorter in higher voltage. Nevertheless the main Andreev peak at
zero-bias will still be a special feature to superconducting superlattices. 
{\em Most importantly, unlike the width of the Andreev peak in the NS
junction, the width of the Andreev peak can be much less than} $\Delta _{0}$%
. This is due to destructive interference of the Andreev reflections from
all the superconductors. As the normal metal diminishes, the separation
between these Andreev peaks increases, and eventually, only the one at the
zero bias survives. At the same time, the width of the Andreev peak extend
to $\Delta _{0}$. In addition to this feature, there are also small
oscillations between these Andreev peaks. These oscillations are also
results of the interference from all the interfaces. They have fixed period
in terms of energy as demonstrated in Fig. 7(c).

It is tempting to make an analogy between the NS superlattice and the
diffraction gratings (each superconductor cell seems to be analogous to a
slit in the grating). However, crucial differences do exist and make the
interference in the NS superlattice system more complicated: (i) The Andreev
reflection from each superconductor has to pass all other superconductors in
front of it to get to the electrode where the interference happens. (ii) The
incident wave that arrives at each superconductor has to pass all other
superconductors in front of it. As a result, a precise dependence of the
separation and the width on $A$ and $B$ are complicated and can only be
computed numerically. Nevertheless, in a special limit in which $A$ is large
and $B$ is very small, one may disregard $B$ in computing the phase. The
result is similar to that for the diffraction grating: Because the path
between successive superconducting cells is $2AV$, \ we have 
\begin{equation}
|C_{0}|^{2}\propto \left[ \frac{\sin (N_{0}AV)}{\sin (AV)}\right] ^{2}.
\label{diffraction}
\end{equation}
Therefore, the positions of \ the main Andreev peaks are independent of the
number of layers $N_{0}$ and are located at $m\pi /A$ with $m$ being
nonnegative integers. On the other hand, the minima are located at $m\pi
/N_{0}A$ with $m$ being nonnegative integers, and hence the width of the
Andreev peak is $\pi /N_{0}A$. All of these results can be explicitly
checked numerically. Fig. 7(d) is a simple demonstration in which the
calculated $dI/dV$ curve based on Eq.(\ref{diffraction}) is compared with
that based on the exact calculation. One sees that Eq.(\ref{diffraction})
gives excellent results in the lower energy sector. Note that for large $A$,
the Andreev peak gets sharper and becomes more like a zero-bias resonant
conductance peak. This is very different from the other limit when $A$ goes
to zero, where the Andreev peak is a plateau, extending from zero-bias to $%
V=\Delta _{0}$.

We now analyze the IS superlattice. Essentially, the $dI/dV$ curve for the
IS superlattice has a similar behavior as shown in Fig. 7(e) in which an
Andreev peak sits at the zero-bias, {\em with reduced width}. The difference
is that the small oscillations now sit in a rising background. The reason
for this rising behavior is due to that when $V$ increases, the insulators
becomes more transparent so that the conductance increases. The differential
conductance shown Fig.7(e) resembles a recent observed data in a underdoped
BSCCO sample\cite{Chang}. This resemblance suggests that a IS superlattice
is more appropriate to model the c-axis transport in the underdoped regime.

\section{Conclusions and Acknowledgments}

In conclusion, we have investigated the effects of the superlattice
structure on the single particle transport along the c-axis of the high Tc
superconductors. Based on superlattice models that consist of
metals/insulators and d-wave superconductors (NS/IS superlatives), we find
that\ a crucial difference between the NS superlattice and the IS
superlattice lies in the positions of the quasi-particle peaks in the
measurement of density of state. In the large mass anisotropy limit of the
metal for the NS superlattice, the density of state in the low energy
section is still bulk d-wave like except that the position of the
quasi-particle peak is reduced considerably, while for the IS superlattice,
the quasi-particle peak remains at the gap value. The width of the Andreev
peak at zero-bias in the planar junction measurement is also shown to be
affected strongly by the superlattice structure. It is found that this width
can be considerably reduced due to the destructive interference of the
Andreev reflections from all the superconductors. In addition to this
feature, there are also distinct oscillations with smaller amplitudes
extending out from the main Andreev peak. Most importantly, for the IS
superlattice, these oscillations sit in a rising background, in consistent
with experiments.

This work is supported by the NSC of Taiwan under Grant No. NSC
89-2112-M-007-091. We thank Profs. H.H.Lin, T.M. Hong and C. S. Chu for
useful discussions.

\section{FIGURE CAPTIONS}

\vspace{0.3cm}Fig.1 The unit cell in a NS superlattice and the definition of
angle $\phi $. We shall denote $d=a+b$ and assume that the $q_{x}$ and $q_{y}
$ directions are infinite.

Fig.2 (a) The one dimensional density of state for the case when $\theta =0$%
, $\phi =0$, $m=m^{\prime }$ and $m_{N}=m_{S}$. Here $A$ and $B$ are chosen
to be close the data for YBCO: $A=0.57$, $B=0.21$. It is seen that the
density of state possess particle and hole symmetry. (b) The one dimensional
density of state for the same parameters used in (a) except that now $%
m_{S}/m_{N}=0.5$. The asymmetry between particles and holes are evident.

Fig.3 (a) The three dimensional density of state for the homogeneous case
when $A=0$ and $B=1$. Here $n(\varepsilon )$ is measured by the unit $%
k_{F}^{2}/4\pi ^{2}d$. (b) A similar plot for the case when $A=0.1,0.2$ and $%
B=1$. One sees that the quasi-particle peak moves in the subgap region,
leaving a peak-like structure at $\varepsilon =1$. (c) A more realistic case
when $A=0.57$ and $B=0.21$.

Fig.4 The three dimensional density of state in the infinite mass anisotropy
limit ($m/m^{\prime }\rightarrow \infty $, $A=0.57$, and $B=0.21$) for the
low energy section. The position of the quasi-particle is considerably
smaller than $\Delta _{0}$.

Fig.5 The three dimensional density of state for the SI superlattice in the
simplest case when all the masses are the same and the other paramters are $%
A=0.57$, and $B=0.21$. Here $V=1.2\varepsilon _{F}$ and $\varepsilon _{F}=10$%
. One sees that the quasi-particle peak remains at $\varepsilon =1$.

Fig.6 Schematic plot of the setup for low contact resistance measurement.

Fig.7 (a) The calculated differential conductance for a NSN junction based
on Eqs.(\ref{BC}) and (\ref{dIdV}) for different lengths of the
superconductor. Here $\phi =0$, $\varepsilon _{F}=100$, and $dI/dV$ is
measured in terms of $2N(E_{F})ev_{F}{\cal A}$. (b) The $dI/dV$ curve
(integrated over $\phi $) for \ a NS superlattice with 10 layers. Here we
assume $m=m^{\prime }=m_{S}=m_{N}$, $A=0.57$, $B=0.21$, $Z=0$, and $%
\varepsilon _{F}=20$. (c) The energies of local maxima of the $dI/dV$ versus
their indices. Except a few steps in-between, the energy is linear in the
index, indicating that the period of small oscillations in Fig.7(b) is a
constant. The steps occurs when one jumps across the main Andreev peaks. (d)
Solid line: The $dI/dV$ curve for a NS superlattice with 15 layers for $A=1$
and $B=0.01$ . Here we assume that $m=m^{\prime }=m_{S}=m_{N}$, $Z=0$, $\phi
=0$ and $\varepsilon _{F}=20$. Dash line: A fitting based on Eq.(\ref
{diffraction}) gives excellent results in low energies. (e) The differential
conductance for a SI superlattice with 10 layers. The parameters used here
are $A=0.57$, $B=0.21$, $Z=0$, $\gamma _{1}=0.5$, $\gamma _{2}=2.87$, $%
\varepsilon _{F}=20$, $V/\Delta _{0}=22$. This curve is close to to the $%
dI/dV$ curve in low temperature recently observed in Ref.\cite{Chang} (see
their Fig.2(b)).


\begin{references}
\bibitem{Ivchenko}  E.L.Ivchenko, G.E.Pikus, Superlattices and Other
Heterostructures-Symmetry and Optical Phenomena 2nd. (Springer, Germany,
1997).

\bibitem{Oomi}  See, for example, G.Oomi, Y.Uwatoko, T.Sakai, H.Fujimori, J.
Magn. Magn. Mater. {\bf 156}:(1-3) 402-404 (1996); L.Lazar, J.S.Jiang,
G.P.Felcher, A.Inomata, S.D.Bader, J. Magn. Magn. Mater. {\bf 223}, 299 -303
(2001).

\bibitem{Hahn}  A. Hahn, Physica {\bf B 165\&166}, 1065 (1990).

\bibitem{Martinez00}  J.C. Martinez et al. con-mat/002111; M. Varela et al.
Phys. Rev. Lett.{\bf \ 83,} 3936 (1999); Douglas H. Lowndes et al. Phys.
Rev. Lett. {\bf 65}, 1160 (1990).

\bibitem{caxis}  See. for example, K. Takenaka, et al. Phys. Rev. B {\bf 50}%
, 6534 (1994).

\bibitem{cresisitivity}  For a review, see S. L. Cooper and K. E. Gray, in 
{\it Physical Properties of High Temperature Superconductors IV}, ed. by D.
M. Ginsberg (World Scieftic,\ Singapore, 1994).

\bibitem{Dessau}  For recent progress, see Y. -D. Chuang et al.
cont-mat/0107002.

\bibitem{Bulaevskii}  L. N. Bulaevskii and R. Rammal, Phys. Rev. B {\bf 44},
9768 (1991).

\bibitem{Cheng}  Hung-Wen Cheng, Ph.D. thesis, National Tsing Hua Unversity
(2000).

\bibitem{Sinha98}  See Fig.2 in Saion Sinha and K.-W.Ng, Phys. Rev. Lett.%
{\bf 80},1296 (1998).

\bibitem{Chang}  H.S.\ Chang, H.J Lee, and M. Oda, cond-mat/0107354.

\bibitem{Xiang}  T. Xiang and W. N. Hardy, Phys. Rev. B {\bf 63}, 024560
(2001).

\bibitem{Gelder}  A.P. Van Gelder, Phys. Rev. {\bf 181}, 787(1969).

\bibitem{covington00}  The reduction of the quasi-particle peak was observed
in Ref.\cite{Cheng} for In/BSCCO system. However, for measurements on YBCO,
one observed a dip at zero bias, see for example, M. Covington and L. H.
Greene, Phys. Rev. B {\bf 62}, 12440 (2000).

\bibitem{Ning}  G. S. Ning et al. Phys Rev. B{\bf \ 51}, 4641 (1995).

\bibitem{BTK}  G.E.Blonder, M.Tinkham, T.M.Klapwijk, Phys. Rev. B {\bf 25},
4515 (1982).

\bibitem{BTK1}  See Fig.9 in G.E.Blonder and M. Tinkham, Phys.\ Rev. B {\bf %
27}, 112 (1983).
\end{references}
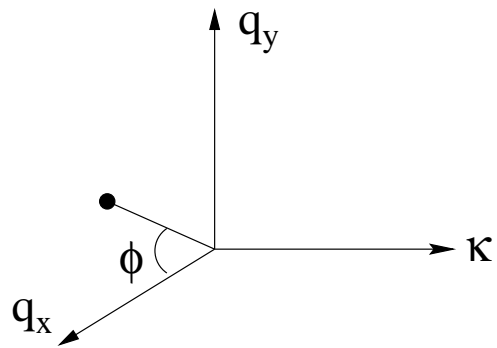
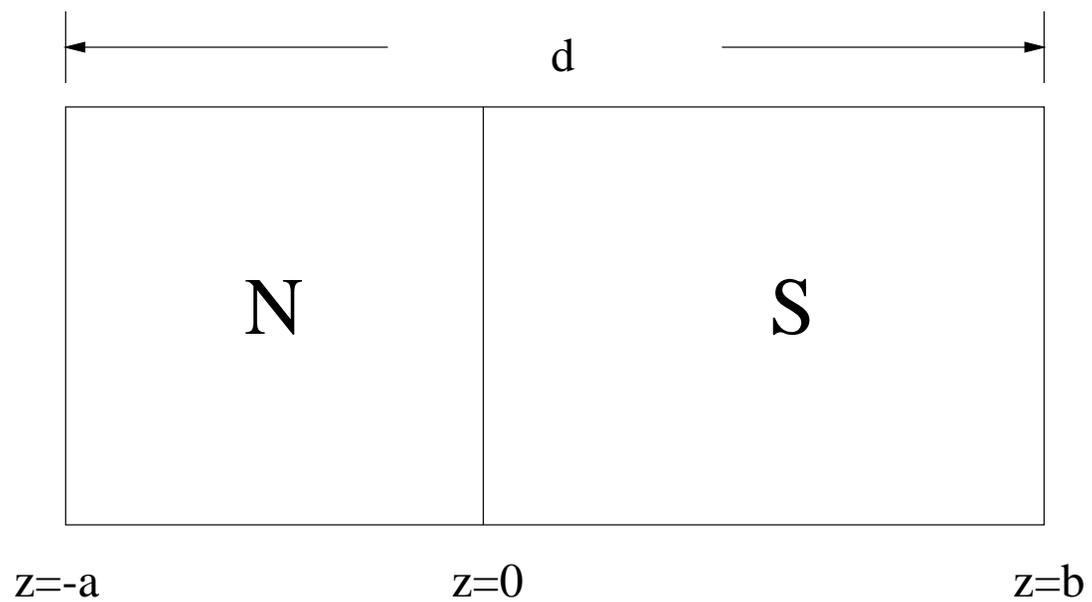
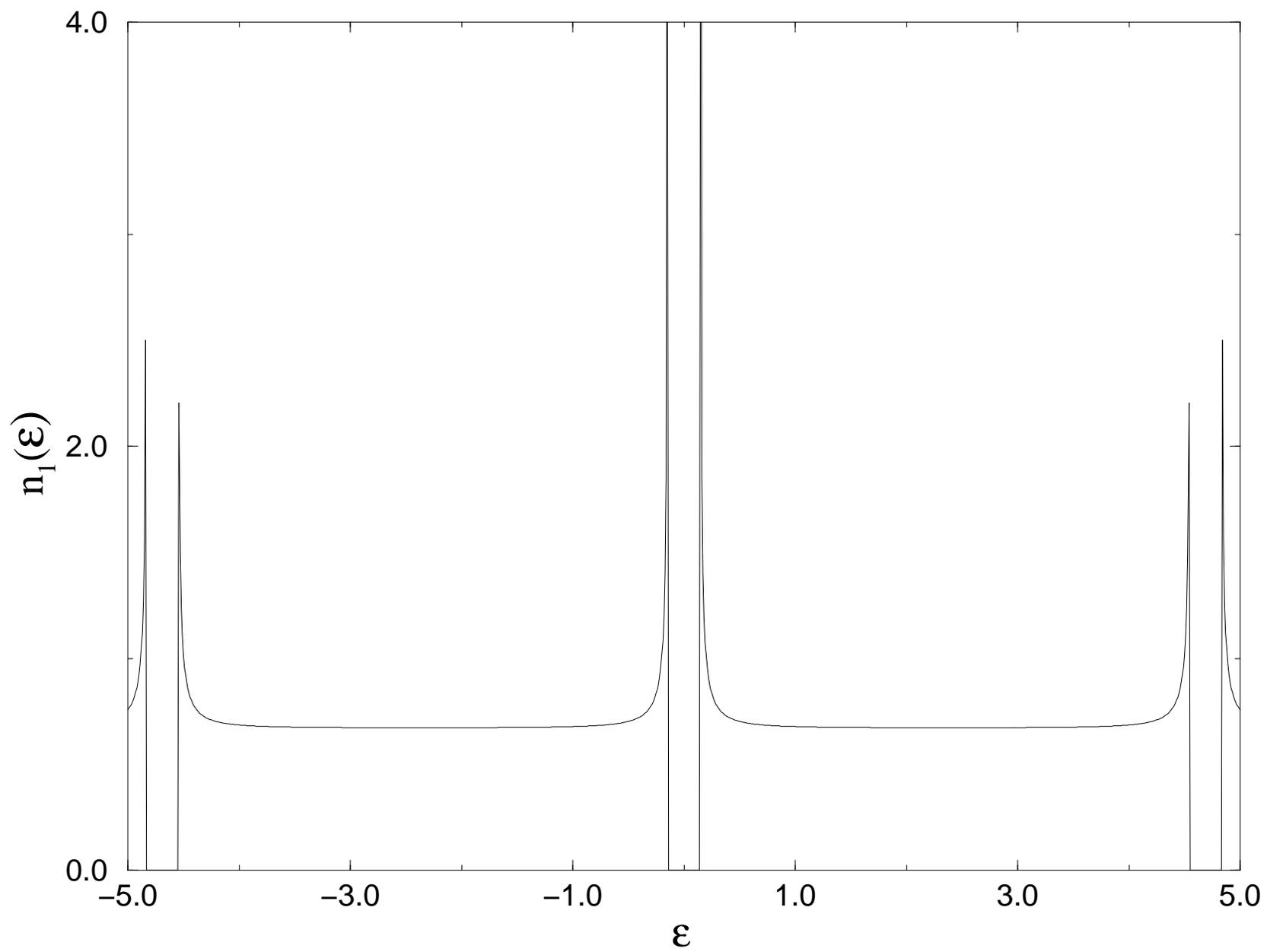
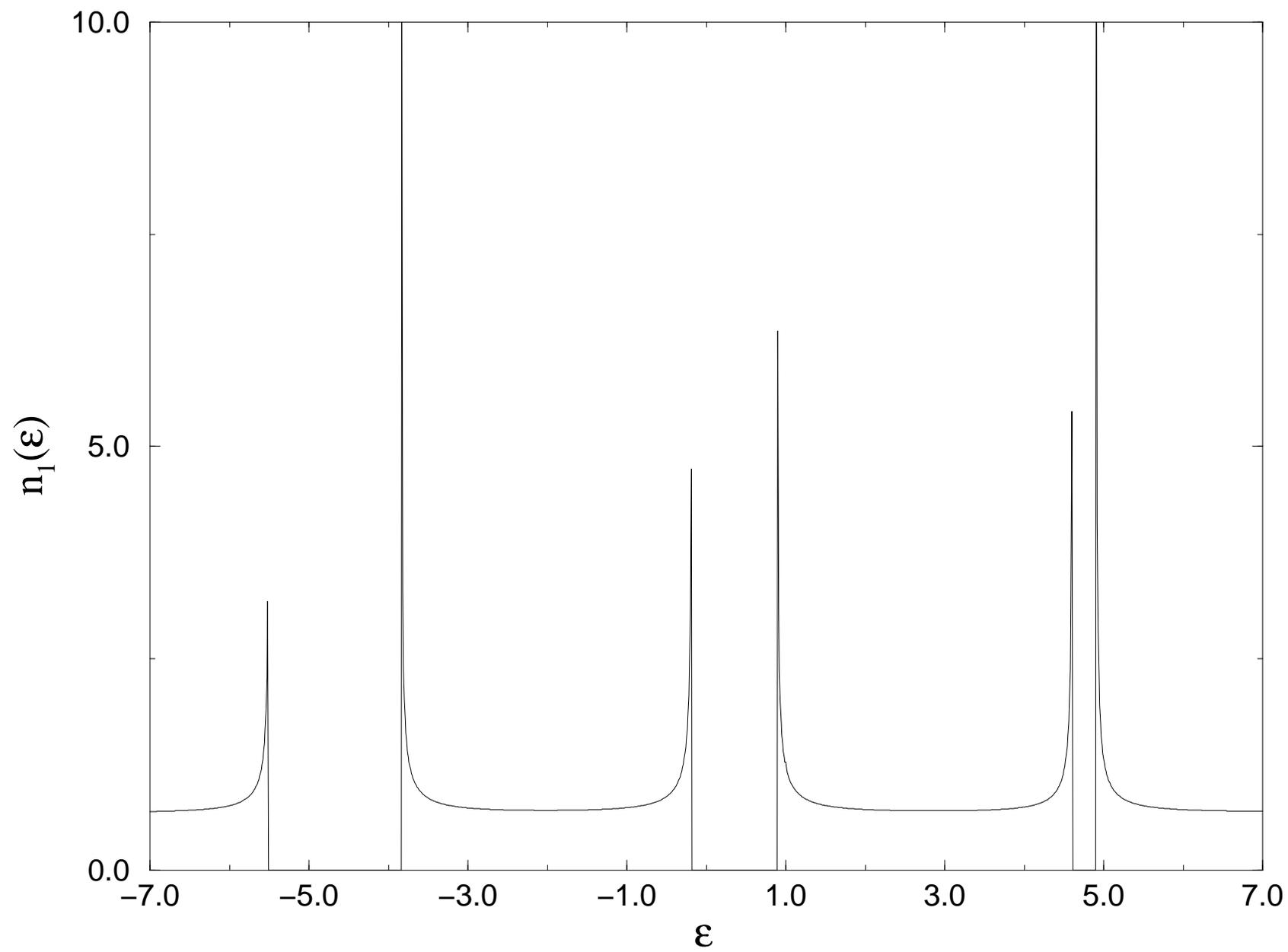
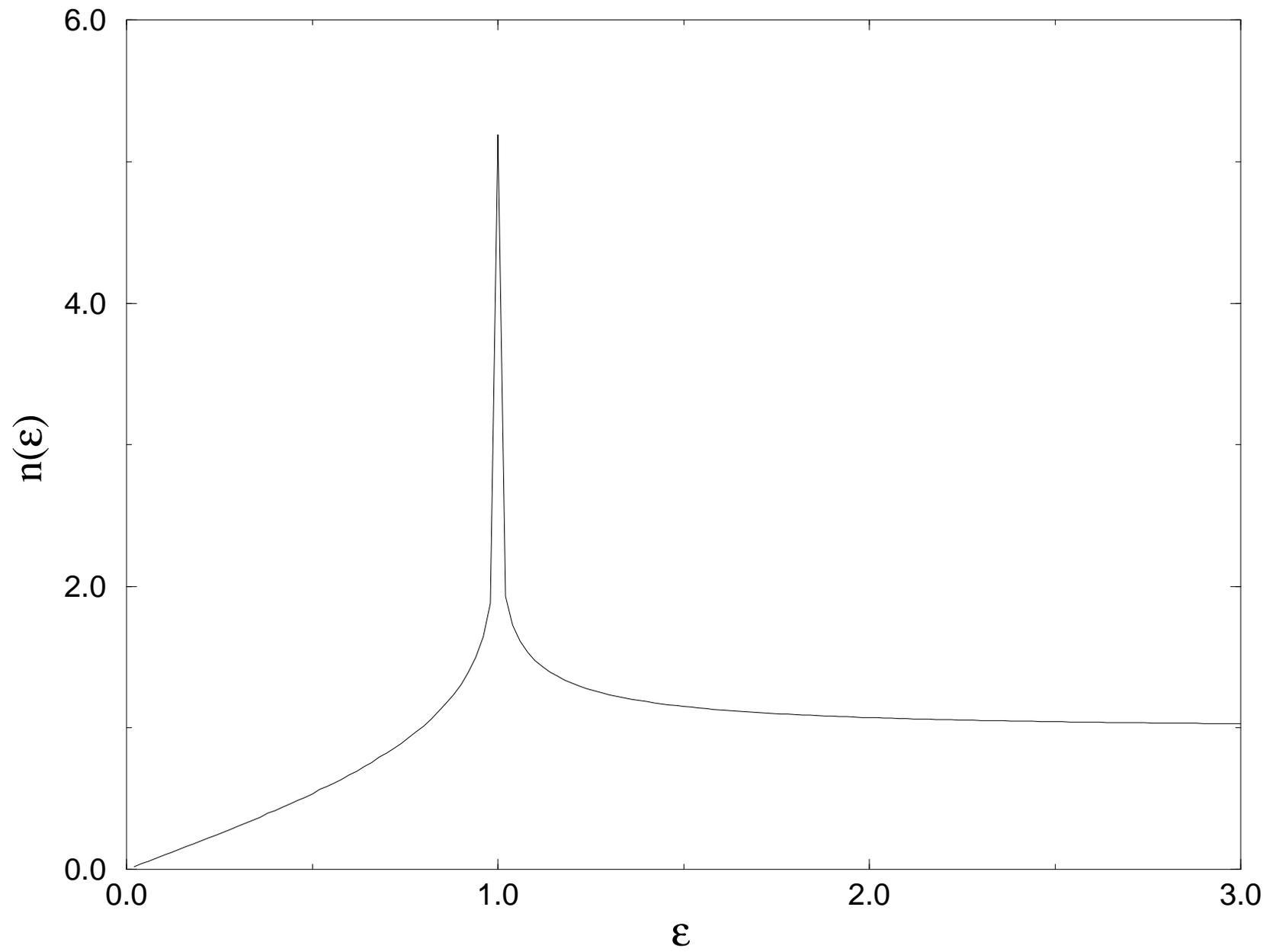
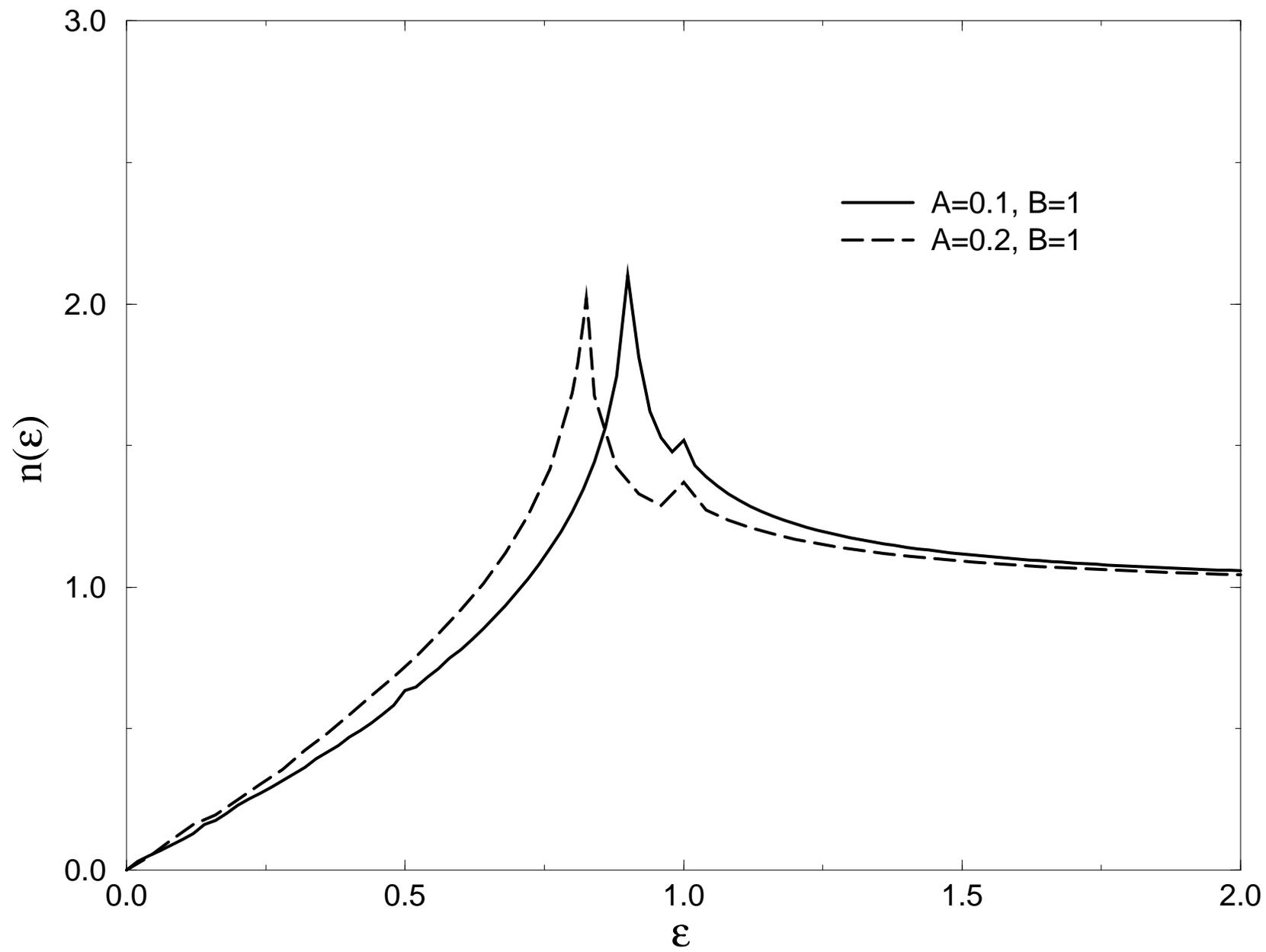
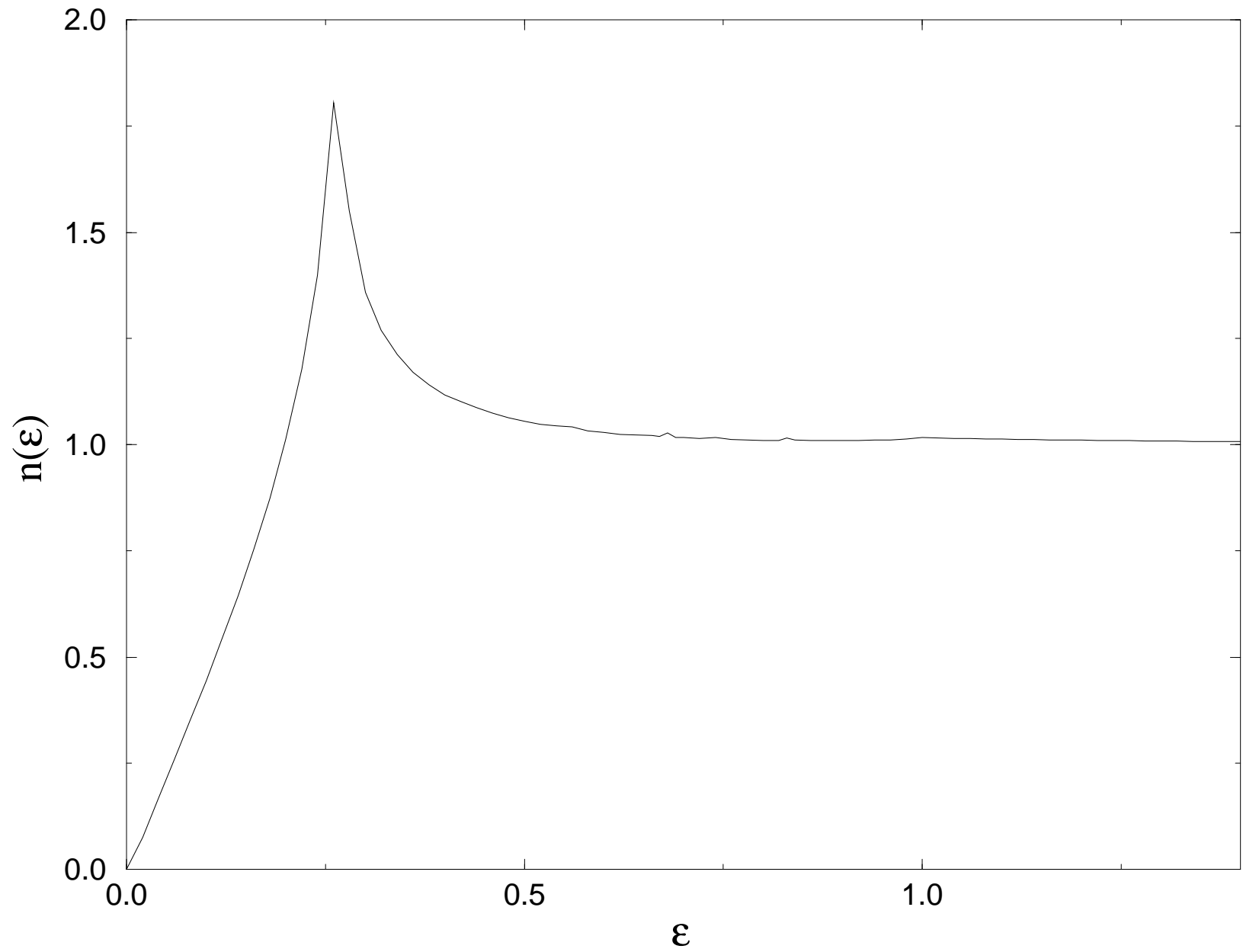
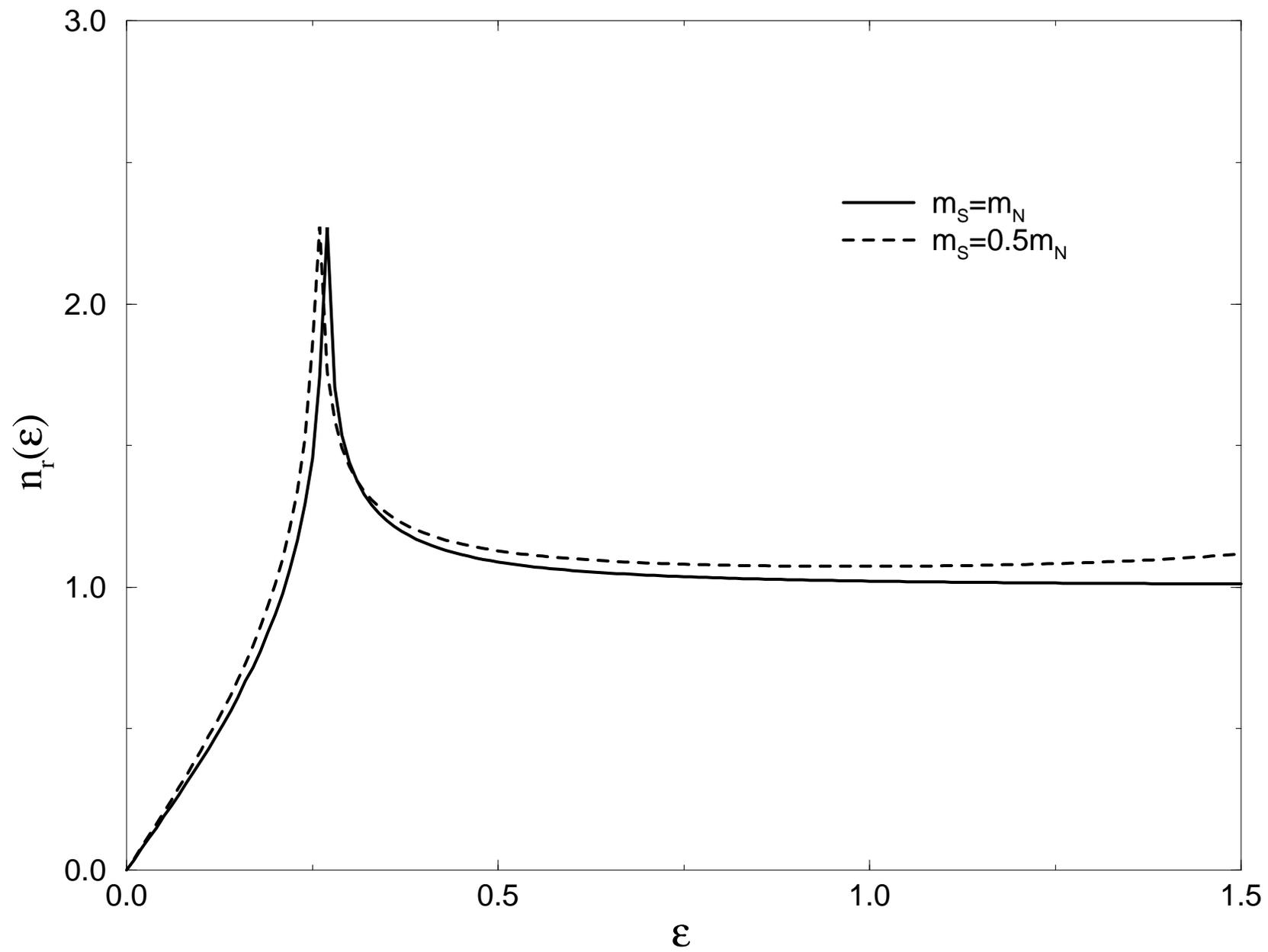
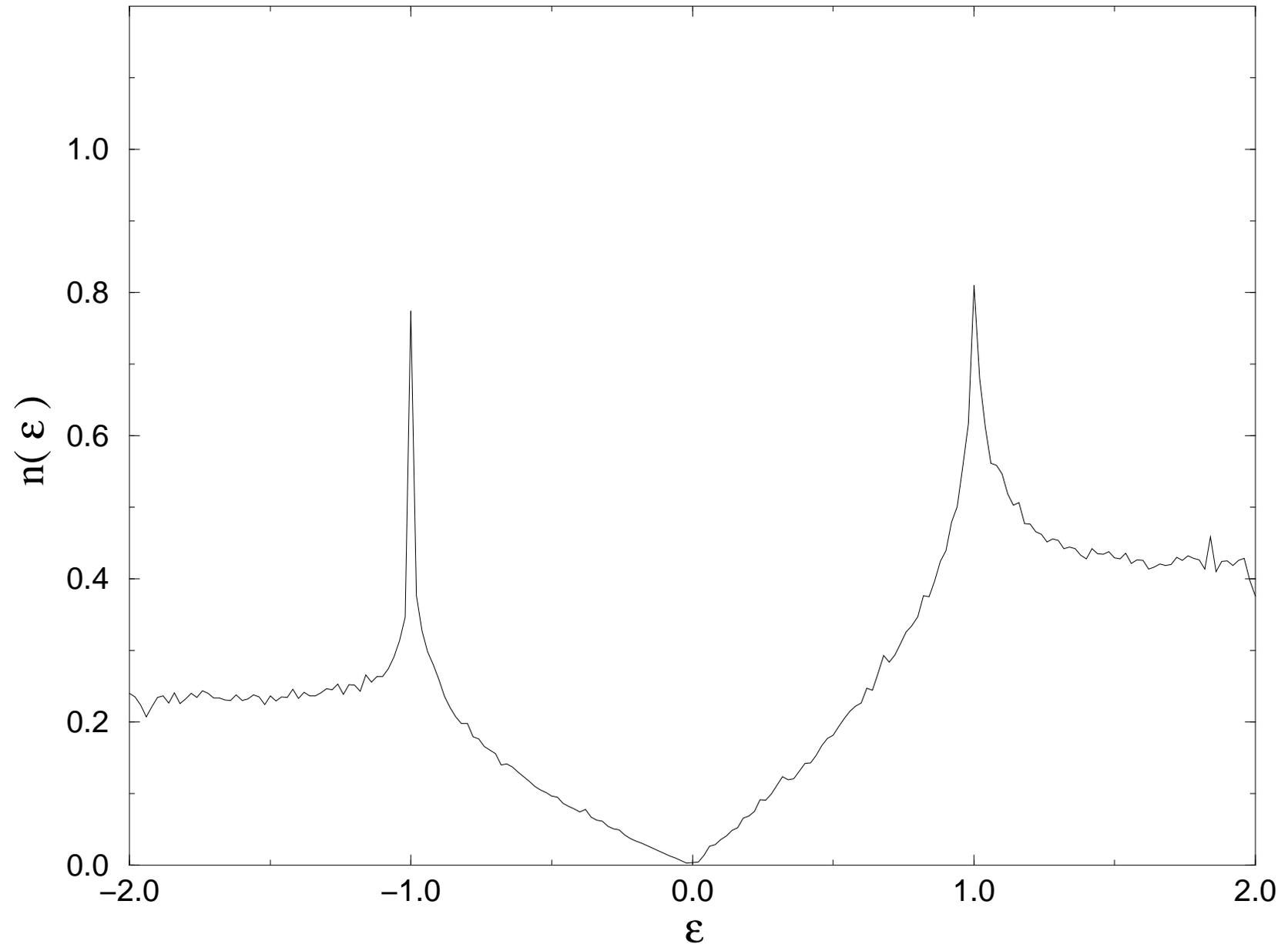
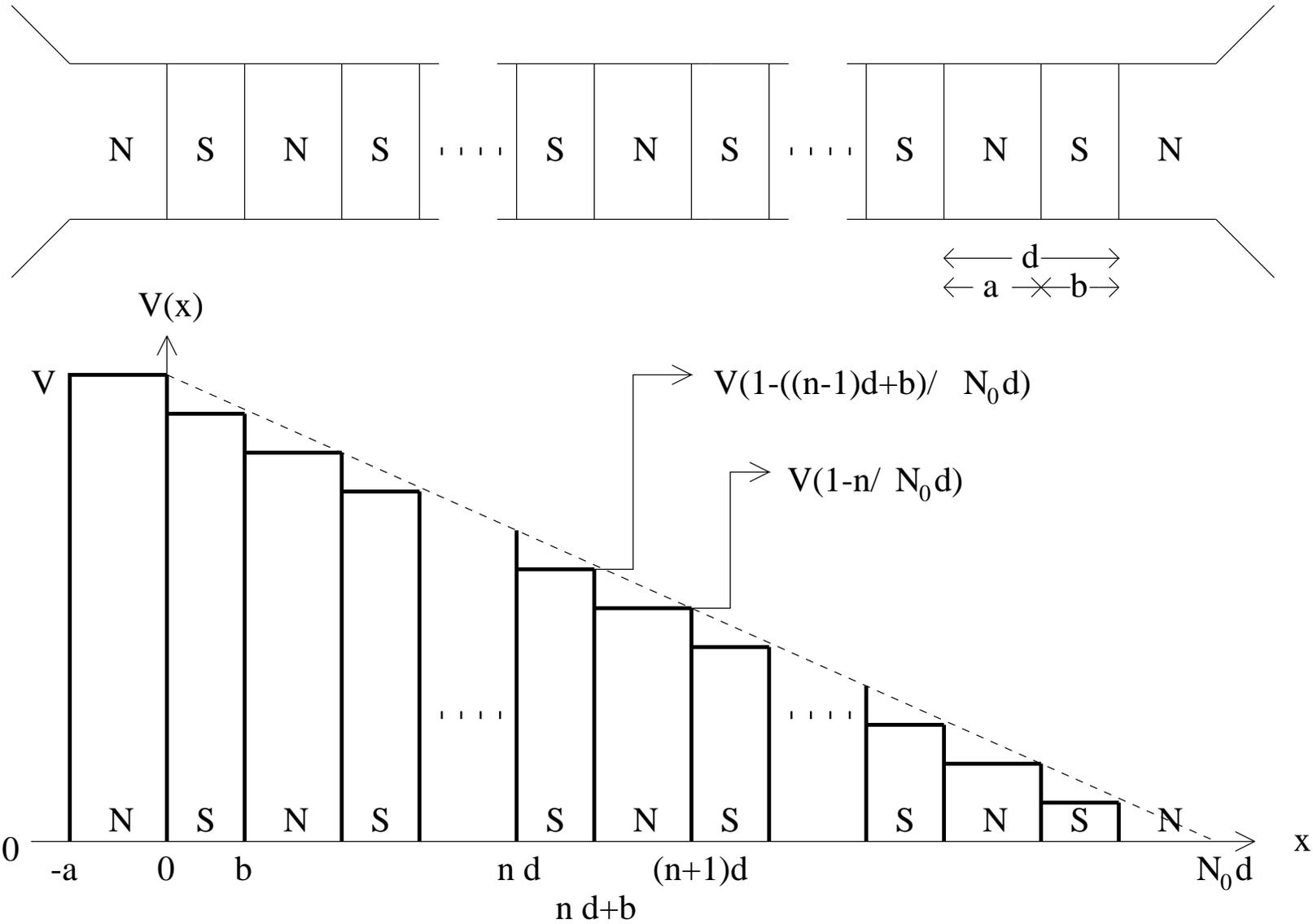
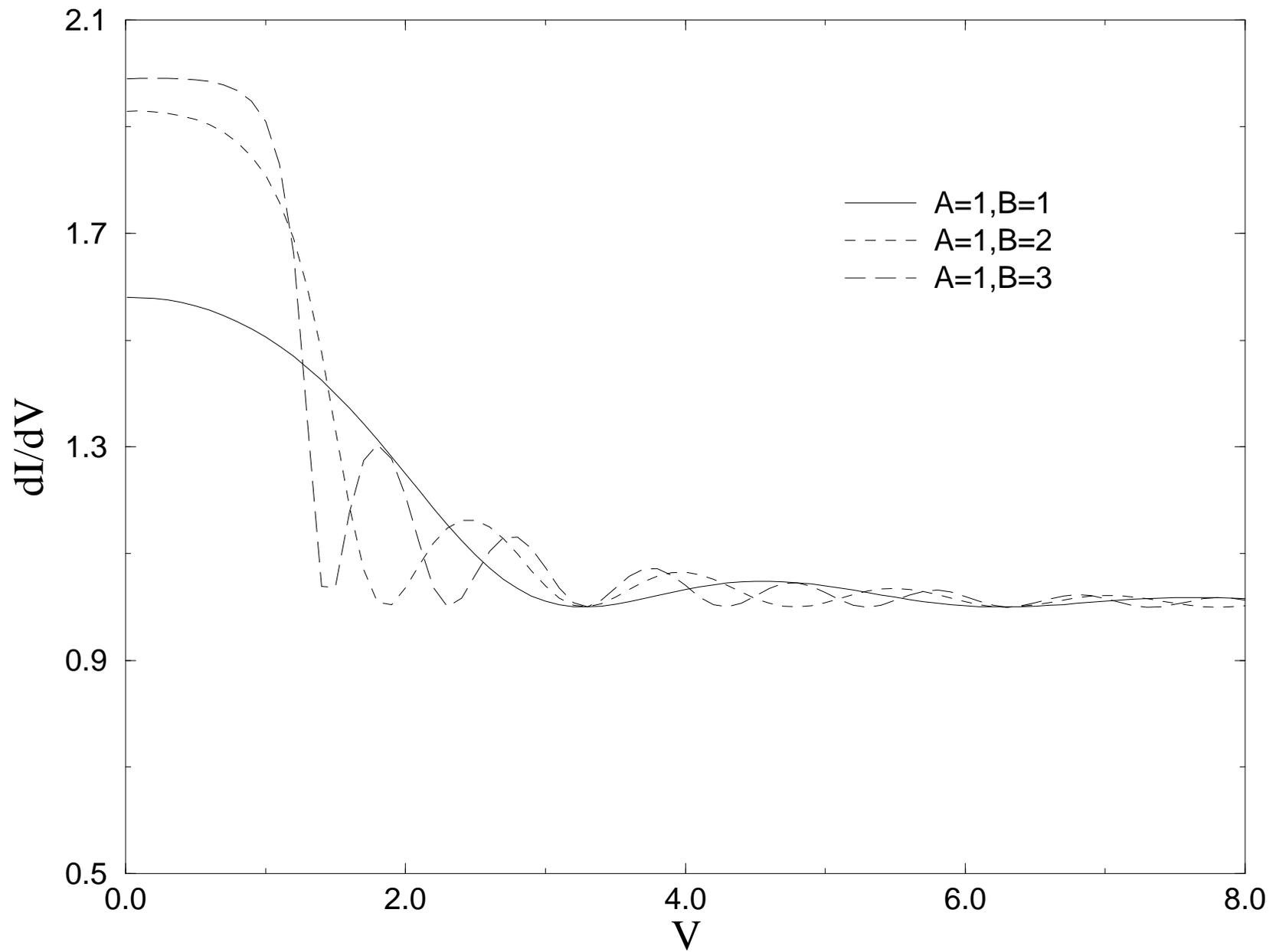
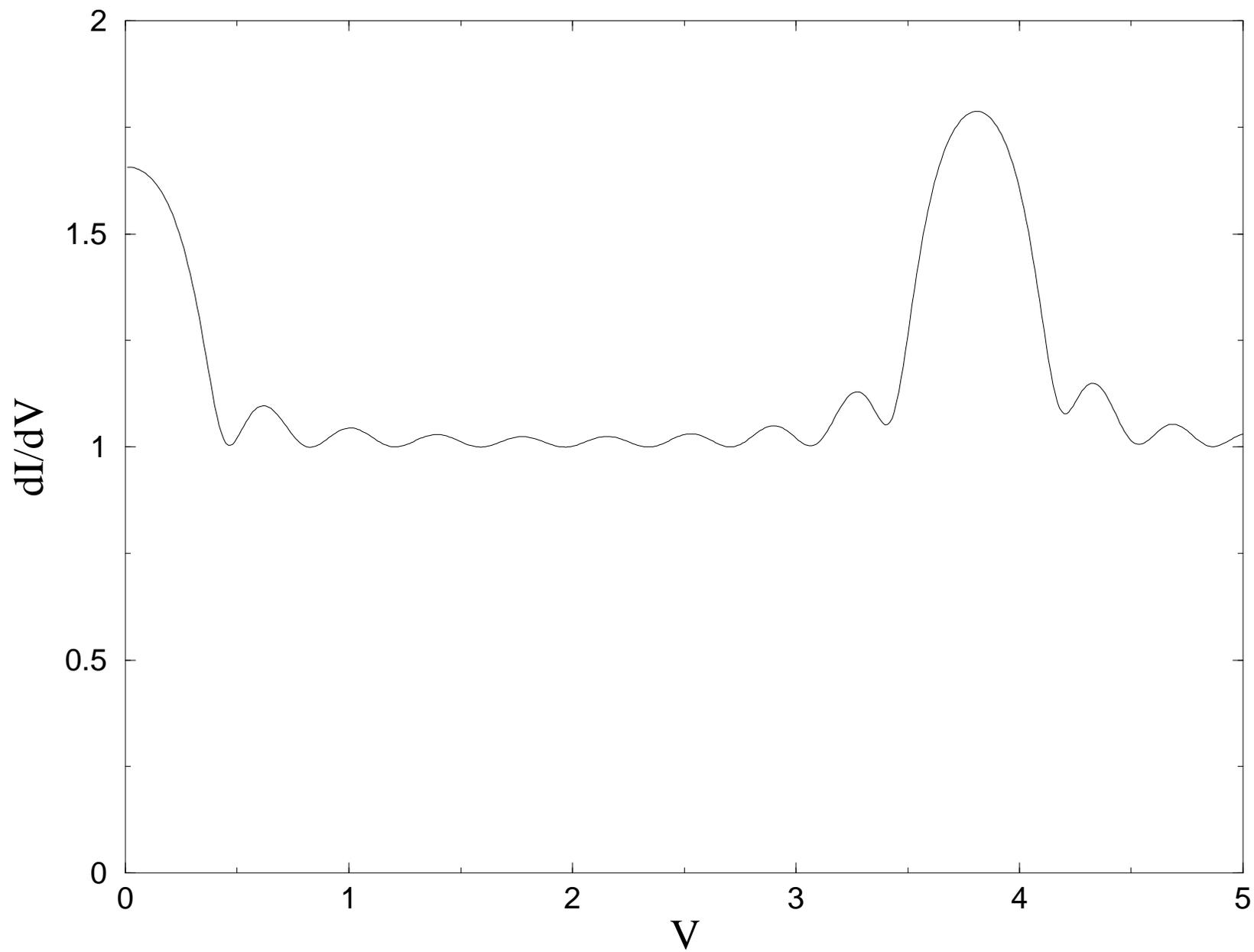
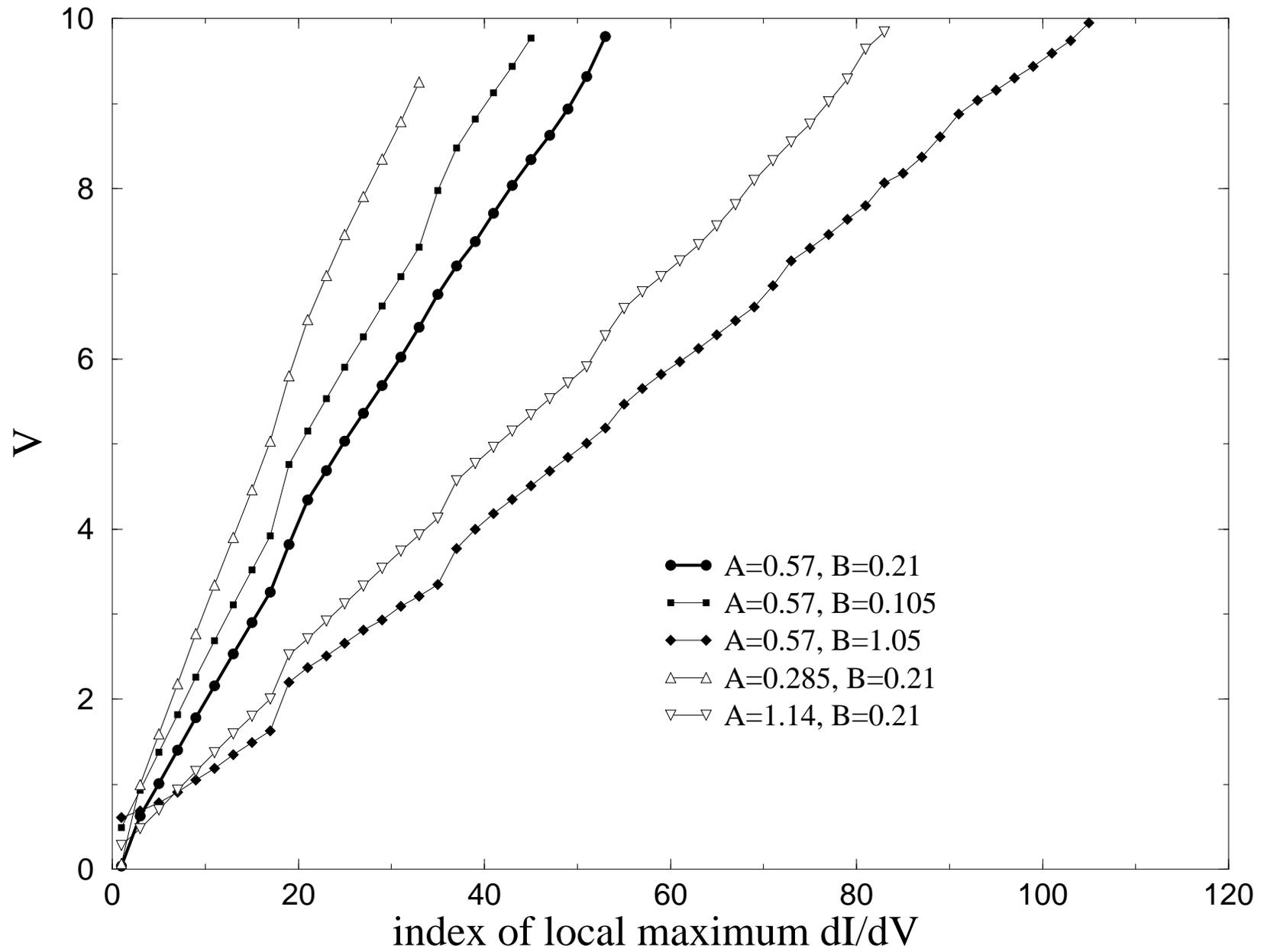
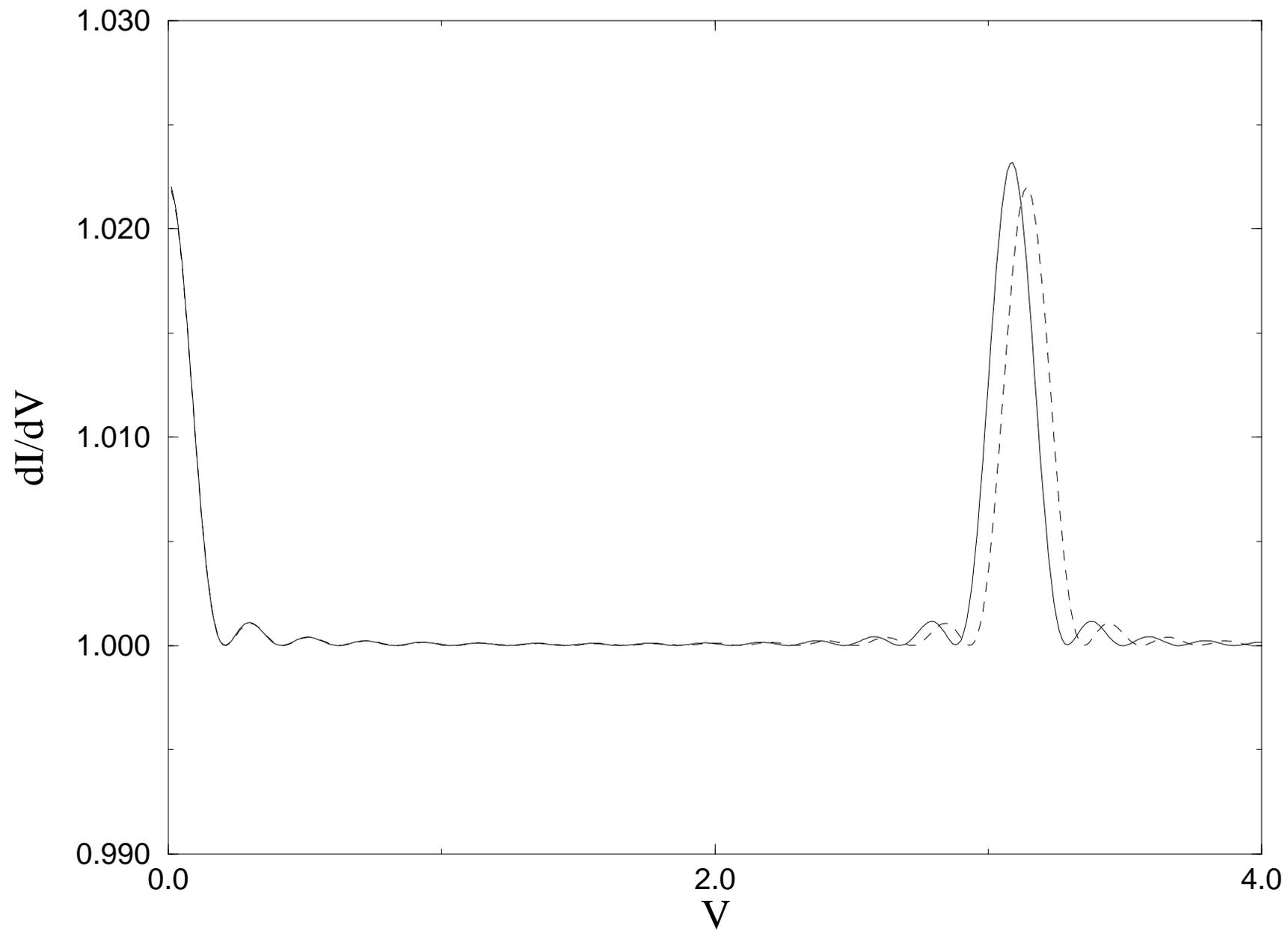
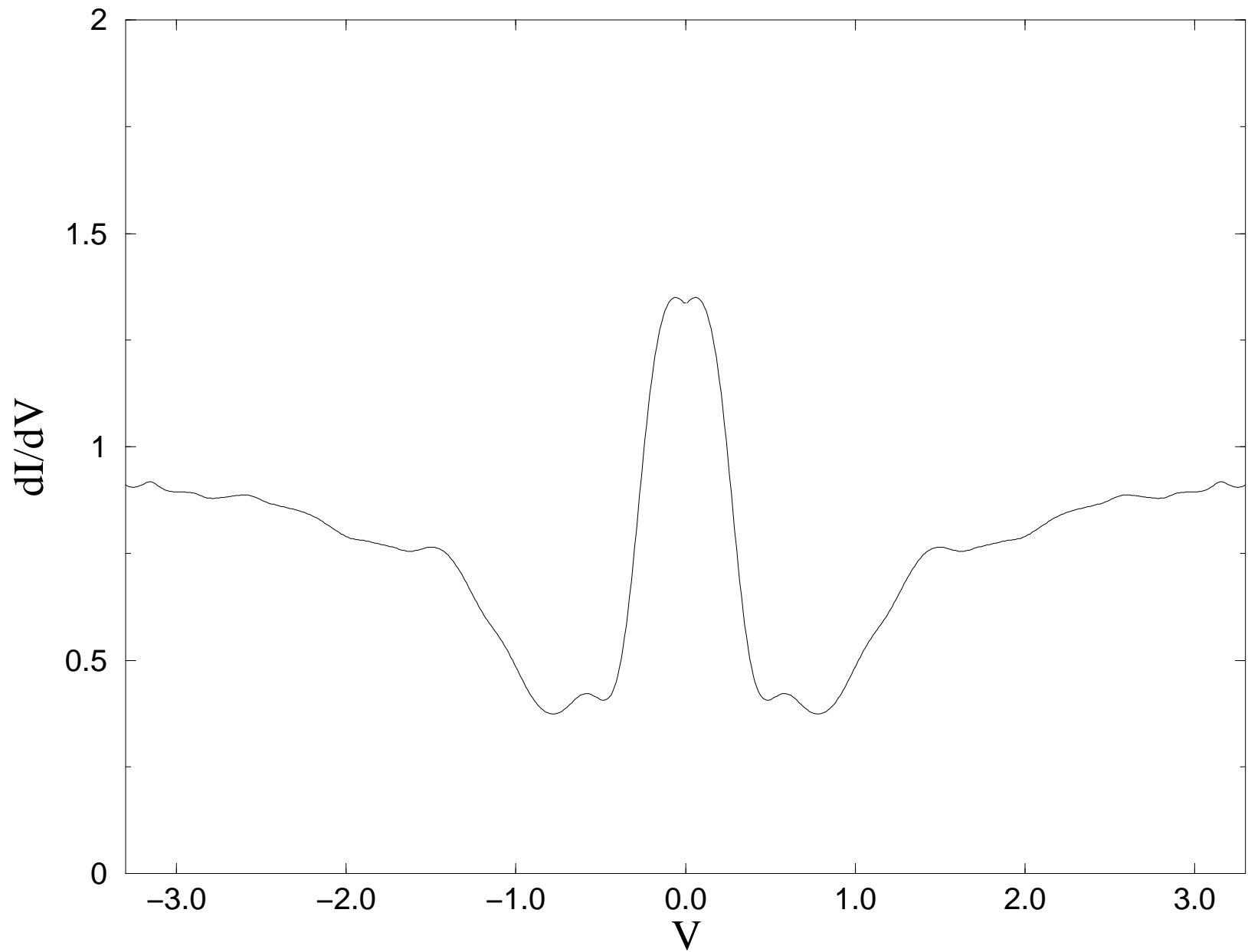
\end{document}